\documentclass[a4paper,12pt,oneside]{article}
\usepackage{graphicx}
\usepackage{amsmath}
\usepackage{amssymb}
\usepackage{euscript}
\begin{document}

\title{
A non-LTE modeling of narrow emission components
of He and Ca lines in optical spectra of CTTS
}

\author{A.V.Dodin, S.A.Lamzin, T.M.Sitnova}

\date{ \it \small
Sternberg Astronomical Institute of Moscow State University,
Universitetskij prospekt 13, Moscow, 119992 Russia
\footnote {Send offprint requests to: A. Dodin e-mail: dodin\_nv@mail.ru}
}

\maketitle

\bigskip



\bigskip


\section*{Abstract}

  A spectrum of a hot spot, produced by radiation of accretion shock at T Tauri
star's surface, has been calculated taking into account non-LTE effects for He\,I,
He\,II, Ca\,I and Ca\,II, using LTE-calculations of spot's atmospheric
structure, calculated by Dodin \& Lamzin (2012).  Assuming that pre-shock gas number
density $N_0$ and its velocity $V_0$ are the same across the accretion
column, we calculated spectra of a system "star + round spot"{} for a set of
$N_0,$ $V_0$ values and parameters, which characterized the star and the
spot.

  It has been shown that theoretical spectra with an appropriate choice of the
parameters reproduce well observed veiling of photospheric absorption lines
in optical band as well as profiles and intensities of so-called narrow
components of He\,II and Ca\,I emission lines in spectra of 9 stars.  We
found that accreted gas density $N_0>10^{12}$ cm$^{-3}$ for all considered
stars except DK Tau.  Observed spectra of 8 stars were succesfully fitted,
asuming solar abundance of calcium, but it appeared possible to fit TW Hya
spectrum only under assumption that calcium abundance in accreted gas was
three times less than solar. We derive spot's parameters by comparison of
theoretical and observed spectra normalised to continuum level, so our
results are independent on unknown value of interstellar extinction.

   We have found that the predicted flux in Ca\,II lines is less than observed one,
but this discrepancy can be resolved if not only high-density but also lower
density gas falls onto the star.  Theoretical equivalent widths as well as
relative intensities of He\,I subordinate lines disagree significantly with
observations, presumbly due to a number of reasons: necessity to take into
account non-LTE thermal structure of upper layers of a hot spot, poorly
known collisional atomic data for He\,I upper levels and inhomogeneity of
the hot spot.


\section*{Introduction}

   Classical T Tauri stars (CTTS) are young (the age $<10^{7}$ yr), low mass
$(M\le 3\,M_\odot)$ pre-main sequence stars, activity of which is caused by
magnetospheric accretion of matter from a protoplanetary disc.  An optical
spectrum of CTTS consists of photospheric spectrum of late type star and
emission lines with non-trivial and variable profiles.  Generally speaking,
CTTS's emission lines consist of two components: a narrow ($FWHM \sim 30$
km\,s$^{-1}$) and a broad $(FWHM > 100$ km\,s$^{-1}$), which are formed in
different spatial regions -- see e.g.  Batalha et al.  (1996), Dodin et al. 
(2012).  Flux ratio of these components is different for different lines,
varies from star to star and with time for the same line.

   It has been known since Joy (1949) that depths and equivalent widths of
photospheric lines in T Tauri stars spectra are smaller than those of
main-sequence stars of the same spectral type.  This effect is used to
explain as the result of superposition of continuum emission onto
photospheric spectrum.  The degree of veiling of CTTS's photospheric
absorption lines in some spectral band can be characterized by value 
\begin{equation}
r = {EW_0 \over EW}-1,
\label{er-definit}
\end{equation}
averaged over lines of that regin, where $EW$ and $EW_0$ are equivalent
widths of photospheric lines in spectra of CTTS and template star of the
same spectral type respectively.

  Line and continuum emission observed in CTTS's spectra can be explained as
follows in the frame of magnetospheric accretion model (K{\"o}nigl, 1991;
Lamzin, 1995).  The matter from the inner disk is frozen in the stellar
magnetic field lines and slides down along them toward the star, being
accelerated by gravity.  Having reached the dense layers of the stellar
atmosphere, matter is decelerated in the accretion shock (AS), converting
into the heat the main part of its kinetic energy, the flux of which is
equal to
\begin{equation}
F_{ac} = {\mu m_p N_0 V_0^3 \over 2},
\label{Fac}
\end{equation}
where $N_0$, $V_0$ are pre-shock gas number density and velocity,
$\mu=1.3$ is the average molecular weight, $m_p$ is the proton mass.  The
hot postshock matter cools down gradually radiating its thermal energy
in the UV and X-ray spectral bands and settles down to the stellar surface.

  One half of the short-wavelength radiation flux of the shock from the
cooling zone irradiates the star, producing so-called hot spot on its
surface, and the second half escapes upward, heating and ionizing the
pre-shock gas.  Calculations show, that pre-shock gas temperature does not
exceed 20\,000 K at a typical velocity $V_0\sim 300$ km\,s$^{-1}$, however
there are ions up to O$^{+5}$ in this region (Lamzin, 1998).

  Assuming that CTTS's emission continuum is formed in the hot spot, Calvet
and Gullbring (1998) have calculated the structure of the atmosphere being
irradiated by the short-wave radiation of the accretion shock without any
allowance for the emission in lines.  However it was discovered later that
photospheric lines in spectra of some CTTSs were filled in by the narrow
emission lines (Petrov et al., 2001; Gham et al., 2008; Petrov et al.,
2011), which were apparently formed in the hot spot and which in addition to
continuum reduce the depth of a photospheric lines.  This phenomenon leads
for example to a variability of the radial velocity of photospheric lines
and/or to a non-monotonic spectral energy distribution (SED) of the veiling
continuum (Stempels and Piskunov, 2003), if to derive it by means of
Eq.(\ref{er-definit}).

  The calculations of the thermal structure and SED of the hot spot with
allowance for the lines as well as the continuum for the first time have
been performed by Dodin and Lamzin (2012) in LTE-approach. The calculations
were performed for various sets of a parameters characterizing the star
($T_{ef},$ $\log\,g$) and an accretion shock ($N_0,$ $V_0$) with solar 
element abundance.

  It was assumed that there was one circular spot at stellar surface, within
which $V_0$ and $N_0$ were constant.  We calculated how the resulting spectra
of a star with $T_{ef}$ from 3750 to 5000 K $+$ a spot should looks like at
different relative sizes of the spot $f$ and its positions relative to the
Earth, characterized by the angle $\alpha$ between the line of sight and
spot's axis of symmetry.  For each of the stars in which veiling by lines
was found (Gahm et al.  2008; Petrov et al.  2011), we were able to select a
model with a spectrum similar to the observed one, at least in the sense
that the lines exhibiting an emission feature in the observed spectra also
exhibit emission features in the models.  Thus, it was shown that the most
strong emission lines being formed in the spot appeared in the spectra of
CTTS as narrow components of emission lines, while more weak emission lines
decreased the depth of respective photospheric lines.

  However a quantitative comparison of simulated and observed spectra
revealed a number of problems: there were not emission lines in some CTTS's
spectra predicted by the models; theoretical spectra did not reproduce
observed intensity ratio of He\,I and He\,II lines; the predicted intensity
of Ca\,I emission lines was so much greater than observed that Dodin and
Lamzin (2012) suggested a substantial depletion of Ca in the accreted gas. 
However, the authors supposed that assumptions taken in the modeling were
more probable reason of the discrepancy.  First of all it is LTE-approach,
because, as noted Sakhibullin (1997), the intensity of emission lines is
usually smaller in non-LTE spectra.

  It will be shown below that if to derive level populations and 
ionization degree of He and Ca from statistical equilibrium equations
instead of Boltzmann and Saha equations, then the agreement between the theory
and observations becames much better for a number of cases.

  We, as well as Dodin and Lamzin (2012), consider only narrow components of
emission lines, which are formed in the post-shock zone, where a problem of
radiative transfer can be treated in the plane-parallel approximation. 
Broad components of CTTS's emission lines in one way or another are formed
in several regions: in the pre-shock H\,II zone, in the region near
truncation radius, where disk matter is freezes in magnetic field lines of
the star as well as in CTTS's outflow.  Hydrogen lines are example of the
lines, in which the broad component usually dominates.  A simulation of an
intensity and profiles of emission lines broad component is much more
complicated problem, which suggests simultaneous solution of 3-D MHD
equations and 3-D radiative transfer -- see, for instance, Kurosawa and
Romanova (2012) and references therein.


\section{The method of non-LTE calculations of He\,I and He\,II level populations }

  The software for the calculations of helium ionization degree and level
populations was developed by ourself. The selection of atomic levels
(so-called the atomic model) that we used in our calculations is described in
the Appendix.

  The region behind the shock front can be divided into two regions for
convenience: the post-shock cooling zone, which is transparent to X-ray and
ultraviolet radiations of the cooling gas, and the hot spot, i.e.  the
stellar atmosphere, where this radiation is absorbed.  Distributions of the
temperature $T$ and the number density over the post-shock cooling zone were
taken from Lamzin's paper (1998) for the parameters $V_0=200-400$
km\,s$^{-1}$ and $\log\,N_0=11.5-13.0.$ The distributions of the same
parameters in the hot spot for the same values of $V_0,$ $N_0$ and for
various parameters of an undisturbed atmosphere of the star were calculated
by Dodin and Lamzin (2012).

\begin{figure}
 \begin{center}
  \includegraphics[scale=0.5]{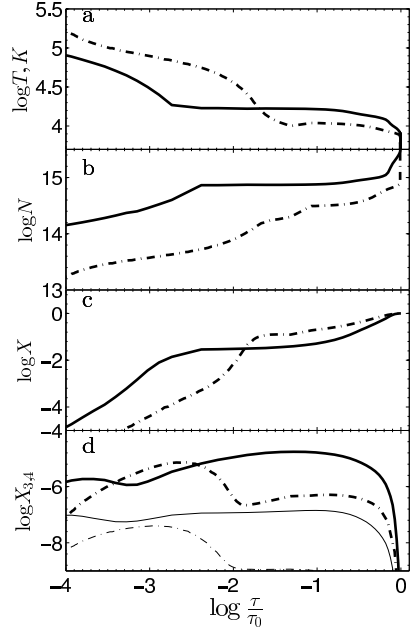}\\
  \caption{
  The distribution of the parameters in the formation region of the He\,II
lines for a star with $T_{ef}=4000$~K and $\log g =4.0:$ the temperature
(a), the number density (b), the relative abundance of He$^+$ (c), the
relative abundance of the levels of He\,II with $n=3$ (d, thick lines) and
$n=4$ (d, thin lines).  The solid lines correspond to the model with
$V_0=400$ km\,s$^{-1}$, $\log N_0=12.5$ ($K=7.6$), the dash-dotted line
corresponds to the model with $V_0=200$ km\,s$^{-1}$, $\log N_0=12$
($K=0.3$).  $\tau$ is the optical depth of the line at $304$ {\AA} with the
maximum value $\tau_0,$ which correspondingly equals to $3.0\times10^4$ and
$1.0\times10^4$.  } 
\label{structure}
\end{center}
\end{figure}

  The spectrum of the hot spot is mainly determined by the parameter $K$
(Calvet and Gullbring, 1998; Dodin and Lamzin, 2012), which equals to the
ratio of the flux $F_{ac}$ of the accretion energy -- see Eq.(\ref{Fac})
-- to the stellar flux:
$$
K = \frac{1}{2}{ F_{ac} \over \sigma T_{ef}^4 }.
$$
The distributions of $T,$ $N,$ and He$^+$ ion relative abundance as well as
the populations of the levels with the principal quantum number $n=3$ and
$n=4$ for two models with a difference in $K$ about 25 times are shown in
Fig\,\ref{structure}.

 The choice of the ion and the principal quantum numbers for the figure is
due to the fact that He\,II 4686 {\AA} line will be actively used to determine 
accretion flow parameters.  In order to display on the figure only the regions
with a noticeable abundance of He\,II, we use a logarithm of He\,II 304
line's optical depth normalized to its maximum value as the abscissa.

%
%

  The system of equations describing a balance of the level populations in a
multilevel atom in stationary conditions is well known (see, for instance,
Mihalas, 1978; Sakhibullin, 1997) and it is useless to write it here.  We will
describe only which atomic data were used and how the system of equations
was solved.

 The Einstein coefficients and the oscillator strengths for all (213 in
total) transitions in our atomic model of He\,I were taken from the NORAD
database (Nahar, 2010).  The same values for He$^+$ ion were calculated by
means of formulae adopted from the textbook of Berestetskij et al.  (1989). 
He\,I recombination coefficients were taken from the NORAD database,
and the same for He\,II -- from the Cloudy08 program (Ferland et al.,
1998).  The photoionization cross section for each level of He\,I was taken
from the NORAD database, and for He\,II it was calculated according to
Golovatyj et al. (1997).

  The rate coefficients for electron impact excitation and de-excitation for
He\,I and He\,II were calculated by means of data from the CHIANTI v.5.0 database
(Dere et al., 1997; Landi et al., 2006) for each transition from $n=1,$ 2
levels to levels with $n\leqslant 5$. The impact rate coefficient
for the $n=3\rightarrow 4$ transition of He\,II was calculated according to
Sobel'man et al. (2002), and the same values for the rest transitions of He\,I 
and He\,II according to van Regemorter (1962).

   The rate coefficients for electron impact ionization from $n=1$ and 2
levels of He\,I were calculated in Born approximation in the book of
Sobel'man et al.  (2002).  For the rest levels of He\,I and He\,II we
consider the electron impact ionization by using the cross section for the
Hydrogenlike atom according to Clark et al.  (1991).

   To calculate the level populations one needs to know radiation field in
the continuum, which is characterized by the mean intensity $J_{\nu}(z),$ where
$z$ is the geometrical depth.  The distribution of $J_{\nu}(z)$ in hot spots
was adopted from the calculations performed by Dodin and Lamzin (2012), and
in the cooling zone $J_{\nu}$ was assumed to be equal to $J_{\nu}$ in the
upper cell of respective hot spot's model because this zone is transparent
for the continuum (Lamzin, 1998).

  Radiative transfer in spectral lines at the stage of level population's
calculations was taken into account by the escape probability method, which
supposed to replace the Einstein coefficient $A_{ik}$ for the spontaneous
transition in the equation of the statistical equilibrium to the
$P_{ik}\cdot A_{ik}$ production, where $P_{ik}$ is escape probability of a
photon from the slab.

 The quantity $P$ is a function of line's optical depth and was calculated
in the same manner as in the Cloudy08 code.  Refering to Ferland et al. 
(1998) for details we note only the following.  In our work, as well as in
the Cloudy08, the incomplete redistribution was adopted for all Lyman lines
of the He$^+$ ion.  The special case was provided for the $L_{\alpha}$ line,
the escape probability for which depends also on the continuum
opacity $\kappa_c$ at 304 \AA.  The dependence of $\kappa_c (z)$ for the hot
spot was adopted from the paper by Dodin and Lamzin (2012), and for the
cooling zone was assumed to be equal to $\kappa_c^0 \times N(z)/N_u,$ where
$\kappa_c^0$ is the opacity coefficient in the upper cell of the hot spot
and $N(z)$ and $N_u$ are correspondingly the number density in the cooling
zone and in the upper cell.  The complete frequency redistribution over the
whole profile was assumed for the rest spectral lines.

   Velocity gradients were not taken into account when we calculated line's
optical depth $\tau$ because in regions with $\tau>1$ helium atoms
thermal velocity is a few times larger than the velocity of the gas
inflow, including the region of He\,II resonance lines formation.  It was
assumed in the process of level population calculations that the local
profile of line's absorption coefficient is thermal with the damping wings.

  Generally speaking the level populations of He\,II can be influenced by a
radiation in hydrogen lines, wavelengths of some of which are nearly coincide 
with that of He\,II lines. However it appeared that this effect is not important 
in our conditions. Due to this reason we will not describe how the level populations of
hydrogen were calculated and note only that the method was the same as in
Cloudy08 code and similar to that we used for helium.
 
  Having defined the atomic data and having calculated the rates of all
respective processes in each knot of the grid over $z$-coordinate, we solved
the system of equations, which describe the statistical equilibrium of the
level populations with fixed the electron number density $N_e$ and the
temperature $T.$ During the iteration process we find the solution of the
statistical equilibrium equations and the escape probabilities, which depend
on this solution.  The iterations were repeated until relative difference of
the optical depth $\tau$ between consecutive iterations becames $\geqslant
10^{-3}$.

  To test the program, we compared our results with results calculated with
Cloudy08 for a model of a gas slab with parameters more or less similar to the
parameters of the gas in the formation region of He\,I and He\,II lines (see
Fig.\,\ref{structure}): the gas temperature $T=30000$~K, total number density
$10^{14.5}$ cm$^{-3}$, the slab thickness $H=10^3$ sm\footnote{ We have
choosen an optically thin in continuum slab, because our program assumes a
given field of an ionizing radiation, but in the Cloudy08 code radiation
flux should be fixed at slab's boundary.}, the radiation field is assumed to
be blackbody with $T=30\,000$~K and the dilution factor $10^{-3}.$

  It follows from the comparison that He\,I level populations differ not
more than two times, probably due to the differences in the adopted atomic
data, mainly rate coefficients for electron impacts.  We computed the same
model with the approximate rate coefficients calculated from van Regemorter
formulae for every level and found the difference with our model in the
level populations of He\,I up to three times.  In the case of He\,II the
difference with Cloudy's results was less than 30\%.  In addition, we
calculated the same model but with the dilution factor of 1: as it was
expected, the calculated populations of He\,I and He\,II practically
coincided with the LTE-populations.  This allows us to conclude that our code
works properly.

  Consider now the processes, which define the level populations of
different levels of He\,I and He\,II in two regions described above. 
Characteristic values of density and temperature in these regions are shown
in Fig.\,\ref{structure}.  Let consider at first the post-shock cooling
zone.

  The ground level of He\,I is equally populated by spontaneous transitions
from upper levels and radiative recombination.  The level is basically
depopulated due to collisional ionization and excitations such as at small
values of $V_0$ and $N_0$ the collisional ionization dominates.  The levels
with $n=2$ are basically populated by collisional processes, but at small
$F_{ac}$ the spontaneous transitions are also important.  The levels are
depopulated by electron collisions except $2^1P$ level, for which the
collisions and spontaneous transitions are equally important.  The
populations of the rest levels of the He\,I are determined by the
collisions.  The most upper levels, combined into the superlevel (see the
Appendix), are populated by collisions and at small $F_{ac}$ also by
radiative recombinations.  The levels are depopulated by collisional
transitions and collisional ionizations.  The ionization balance between
He\,I and He\,II is determined by collisional ionizations and radiative
recombinations.

  The ground level of He\,II is populated by spontaneous transitions and
recombinations and depopulated by collisional processes.  The populations of
levels with $n=2-4$ are equally determined by all processes with small
predominance of collisions.  The levels with $n=2-3$ are depopulated by
spontaneous transitions and for $n=4$ collisions as well as spontaneous
transitions depopulate the level.  The levels with $n=5-6$ are populated by
collisions, but for $n=6$ recombinations become important.  The levels are
depopulated by collisional transitions, ionizations and spontaneous
transitions. 

  Consider now the hot spot. In this region the ground level of He\,I is
controlled by photoionization, recombinations and collisional de-excitations
from upper levels.  The depopulation of the level due to collisional
excitation becomes important at small $V_0$.  Every excited levels of He\,I
is controlled by collisions.  For $2 ^3P$ level the spontaneous transition
to the ground level dominates only at small $F_{ac}.$ Superlevel's
population is determined by collisional transitions and collisional
ionization.

  The ground level of He\,II  is determined, on the one hand, by collisional
and radiative ionization of He\,I, by collisional transitions and by a
recombination of He\,III and, on the other hand, by a recombination into
He\,I and a photoionization into He\,III.  The populations of the second and
third levels are controlled by spontaneous transitions and radiative
recombinations.  The levels are depopulated by collisions and spontaneous
transitions.  The level with $n=4$ is populated by electron collisions and
to a less extend by recombinations.  The level is depopulated by collisions
and to a less etend by spontaneous transitions.  The rest levels are
controlled by electron collisions.  Collisional ionization becomes important
for the depopulation of levels with $n\geqslant 6.$

  The ionization balance between He\,I and He\,II is equally determined by
radiative and collisional processes, but the balance between He\,II and
He\,III is basically determined by radiative processes.

  Hot spot's $T(z)$ dependence that we calculated in LTE approximation
probably overestimate true one at Rosseland optical depths $\log
\tau_{Ross}<-3$ (Dodin and Lamzin, 2012).  However if the level populations
and the ionization degree are basically determined by radiative processes,
which depend weakly on the temperature, one can suppose that any probable
deviations of $T(\tau_{Ross})$ from the true dependence do not influence
significantly on level populations.

  To test this hypothesis we compared ionization degrees as well as He\,I
and He\,II level populations in the models calculated in the usual way with
that of the models, where the temperature is set to be constant at
$\tau_{Ross}<\tau_0$ and equals to $T(\tau_0)$.  The values of $\log \tau_0$
were chosen to be equal to $-3.0,$ $-3.5$ and $-4.0.$ The differences in
populations for He\,I turned out to be large enough.  For instance, at $\log
\tau_0=-3$ the column density of He\,I in various excited states can differ
from the basic model up to 3 times (for the level with $n=2$).  The
difference decreases with increasing of $F_{ac}:$ in the case of the model
with $\log N_0=12.5$, $V_0=400$ km\,s$^{-1}$ the deviation for $n=2$ was
30\,\% only.  At $\log \tau_0=-3.5$ the behavior of the deviations is
similar, but the relative difference was always smaller than 50\%.

  The sensitivity of He\,I upper levels population to hot spot's thermal
structure is caused by the fact that the population is controlled by
electron collisions, rates of which strongly depend on gas temperature. 
Moreover, the collisional cross section for upper $(n \geqslant 3)$ levels
are poorly known.  It is hard to say now how the deviation of $T(\tau)$
dependence from the true one and the uncertainties in the collisional cross
sections influence on the intensity of He\,I subordinate lines.  Due to
these reasons we will not use these lines for diagnostic purposes.

  On the other hand it appeared that sensitivity of He$^+$ ion level
population to thermal structure variations is relatively small: the
differences in the populations between the basic models and models with
various $\tau_0$ do not exceed a few percents.  We emphasize that the
problems with the populations of excited levels of He\,I do not influence on
the ionization balance of He\,I-He\,II, because in our case more than 99\%
of He\,I atoms are on the ground level $n=1.$ Thus in the case of He\,II we
can carry out the non-LTE calculations with the LTE-structure of the
atmosphere. We will see later that He\,II at 4686 line is very useful for a
diagnostic of physical conditions in CTTS's accretion zone.


\section{Non-LTE calculations of Ca\,I and Ca\,II level population}

  Calcium lines in CTTS optical spectra belong to Ca\,I and Ca\,II only. 
Due to this reason we will consider here only the hot spot and an
undisturbed stellar atmosphere because Ca ions with the charge $\geqslant
+2$ only exist in noticeable amount in post shock cooling zone.  For Ca\,I
and Ca\,II we will use the LTE thermal structure for an undisturbed
atmosphere, calculated with the ATLAS9 code, and hot spot models described
by Dodin and Lamzin (2012).  Thus we will suppose that changes in Ca level
populations due to the non-LTE effects do not influence substantially on the
opacity coefficient as far as abundance of calcium is small.

  We varied $T_{ef}$ from 4000 to 5000~K, and assumed that $\log\,g = 3.5$
or 4.0.  To reduce the number of free parameters we always assume that the
microturbulence velocity $V_{mic}=2$ km\,s$^{-1}$.

  The model of Ca atom and the method of calculation of calcium level
populations are identical to those in Mashonkina et al.  (2007) paper.  The
only modification was introduced into the DETAIL code (Butler and Giddings)
to take properly into account an external radiation.  More specifically we
changed the boundary condition at $\tau=0$ for $J=\left[I(\mu)+I(-\mu)\right]
/2$ quantity in the \texttt{formal} subroutine, which calculates radiation
field from known source function $S$ and opacity coefficient $\chi.$ Here
$I(\mu)$ is specific intensity of radiation, directed at an angle $\theta$
to the normal of plane-parallel atmosphere, such as $\mu>0$ for an emerging
radiation. The new boundary condition for $J,$ that takes into account an
external radiation with an intensity $I_e$ is (Mihalas, 1978):
$$
\mu\frac{{\rm d}J}{{\rm d}\tau}=J-I_e.
$$
The intensity $I_e$ was calculated by an interpolation of the radiation,
being formed in the cooling zone and calculated by Dodin and Lamzin (2012),
to a frequency grid, which is used in the DETAIL code.

  To verify the modified program we compared the radiation field, computed
by DETAIL, with the field, computed by ATLAS9 (see Dodin and Lamzin, 2012). 
It turned out that for any model at every depth and frequency they coincided
and the main difference was a more precise frequency grid, used in the
DETAIL code.

\begin{figure}[h!]
  \begin{center}
  \includegraphics[scale=0.5]{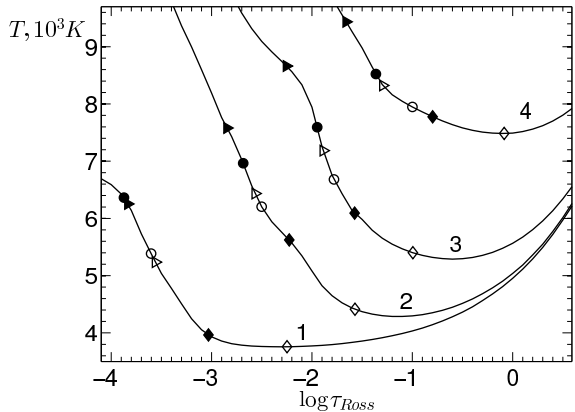}\\
  \caption{The formation region of Ca\,I and Ca\,II lines for solar Ca
abundance (filled-in marks) in the accreted gas and for the abundance reduced
by 10 times (empty marks).  Diamonds correspond to the line of Ca\,I 5589,
rounds correspond to Ca\,I 4226 line, triangles correspond to the
line of Ca\,II 8498.  $T_{ef}=4500$~X, $\log g=4.0$.  For various models:
1 -- $V_0=200$, $\log N_0=11.5$; 2 -- $V_0=400$, $\log N_0=11.5$;
3 -- $V_0=200$, $\log N_0=13.0$; 4 -- $V_0=350$, $\log N_0=13.0.$
See text for details.
}
\label{regionNCLF}
  \end{center}
\end{figure}

  We found that Ca\,I and II lines are originated as a rule in hot spot's
layers with $\log \tau_{Ross}>-3.$ Exceptions are models with small $F_{ac},$
for which the accretion is poorly manifested in the observations -- see
Fig.\,\ref{regionNCLF}.  One can conclude therefore that possible
uncertainty of the temperature in the upper layers of the hot spot cannot
produce significant error in the calculated spectra of Ca\,I and Ca\,II.

%
%

  In the case of the undisturbed atmospheres of CTTS the solar chemical
abundances of all elements, including Ca, were assumed.  However, due to
reasons described in the Introduction we consider the models, where the
calcium abundance in the accreted gas, and therefore in the hot spot, can be
smaller than solar one.  The formation regions of several lines of Ca\,I and
Ca\,II at various calcium abundances are shown in Fig.\,\ref{regionNCLF}. 
It can be seen that for a reduced calcium abundance the line formation
regions are shifted to smaller temperatures, and therefore one can expect 
that the source function at $\tau=1$ as well as the intensity will decrease.

\begin{figure}[t!]
  \begin{center} \includegraphics[scale=0.5]{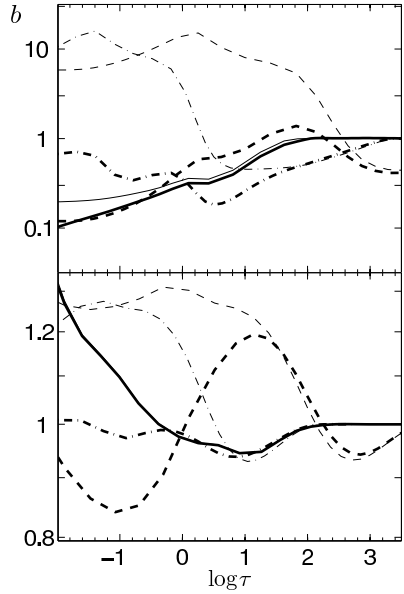}\\ 
 \caption{ The $b$-factors of Ca\,I 4226.7 (dashed), Ca\,I 6162.2
(dash-dotted) and Ca\,II 8498.0 (solid) lines as function of its optical
depth at the line center.  Thick curves correspond to upper levels and thin
curves correspond to lower levels.  In the case of Ca\,II 8498.0 line, the
thick and thin curves are almost merged.  The model of the hot spot was
calculated for stellar parameters: $T_{ef}=4500$~K, $\log g=4.0$ and for
parameters of the accretion flow $V_0,$ $\log\,N_0$: 200, 11.5 (upper
panel); 400, 12.5 (lower panel) 
}
  \label{bf-tau-1}
  \end{center}
\end{figure}

   Character of the deviations from the LTE level populations was different
for different levels and layers of the hot spot: some levels were
overpopulated $(b>1)$ and some were underpopulated $(b<1).$ As an example,
we plotted on Fig\,\ref{bf-tau-1} $b$-factors of levels of some lines, which
will be important in the subsequent diagnostics of CTTS's accretion zone as
a function of the optical depth in the center of each line.  The upper
levels of Ca\,I lines, shown in the figure, are underpopulated or close to LTE
while a lower level is overpopulated, but in the case of Ca\,II 8498 line
departures from LTE are the same for the both levels.  We emphasize that the
departures from LTE became smaller, when infall gas density or its velocity
increase.

%
%

\begin{figure}[t!]
  \begin{center} \includegraphics[scale=0.5]{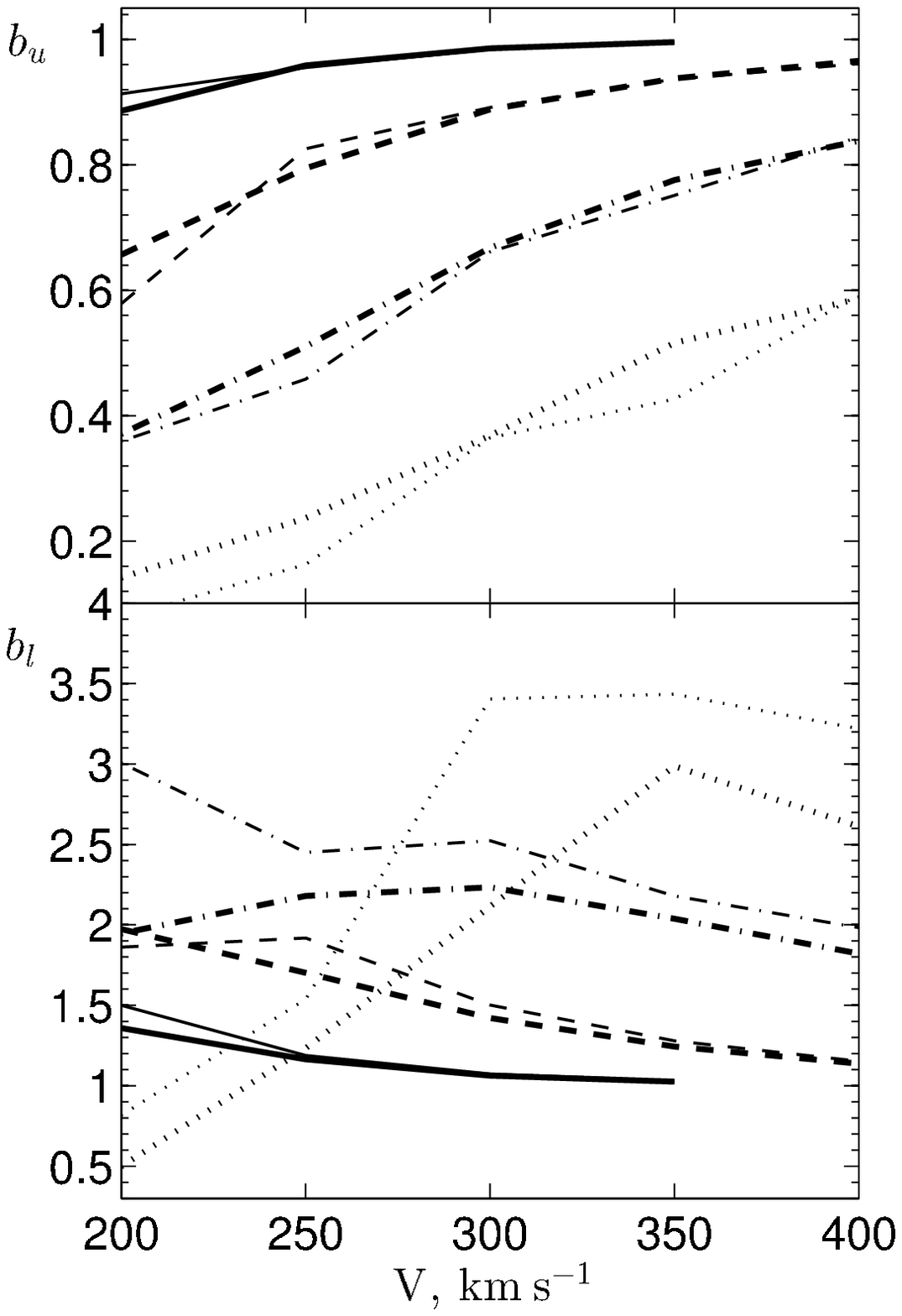}\\ 
 \caption{ The $b$-factors of the upper (top panel) and the lower (bottom
panel) levels of Ca\,I 5589 at the depth, where the optical depth at the
line's center is equals to 1, as a function of the pre-shock gas velocity
$V_0$ (in km\,s$^{-1}$). The solid, dashed, dash-dotted, and dotted lines
are for $\log N_0 = 13.0,$ $12.5,$ $12.0,$ and $11.5,$ respectively.  The
thin and thick curves correspond to the models with the stellar temperature
$T_{ef}=4000$~K and 5000~K respectively.}
 \label{bf-tau-2}
  \end{center}
\end{figure}

  It can be seen more clear on Fig.\,\ref{bf-tau-2}, where we plotted
$b$-factors of upper and low levels of Ca\,I 5589 line at the point of
the hot spot where line's central optical depth is equal to 1 as a function
of accretion parameters $V_0$, $N_0.$ \footnote{The vicinity of this point
gives the main contribution to the central part of line's profile.} The
similar behavior is typical for all considered lines of Ca\,I and Ca\,II.

%
%


\section{ The calculation of spectrum of Ca and other photospheric lines}

  To compute the intensity spectrum of a photosphere, we modified the SYNTHE
code (Kurucz, 1993; Sbordone et al., 2004). The initial code calculates
a LTE spectrum and therefore the source function is assumed to be the same
for all lines and equals to the Planck function: $S_{\nu}=B_{\nu}(T)$.  To
take into account lines with departures from the LTE, we replaced this
expression by the general one:
$$
S_{\nu}=\frac{j_{\nu}}{\kappa_{\nu}}.
$$
In the initial code the factor, which corrects the opacity due to stimulated
emission, does not depend on the specific source of the absorption, however
in non-LTE case the factor may be different for each $i$-th source of
opacity. Therefore an initial expression for the total opacity:

$$
\kappa_{\nu} = (1-e^{-\frac{h\nu}{kT}}) \sum {\kappa^0_i(\nu)}
$$
should be replaced by:
$$
\kappa_{\nu} =
\sum\limits_i \kappa_i(\nu) =
\sum\limits_i \kappa^0_i(\nu)\,
\left[ b_l^i-{b_u^i}e^{-\frac{h\nu}{kT}} \right],
$$
where $\kappa^0_i(\nu)$ is an opacity, concerned with an $i$-th source and
being calculated in the initial code, and $b_u^i$, $b_l^i$ are Menzel's
$b$-factors of upper and low levels for the $i$-th source.

  The expression for the total emission coefficient can be written as:
$$
j_{\nu} = \frac{2h\nu^3}{c^2} \, \sum\limits_i
\kappa_i(\nu) \left[ \frac{b_l^i}{b_u^i} e^{\frac{h\nu}{kT}}-1\right]^{-1}.
$$
To calculate a line in LTE approximation, its $b$-factors should be equated to 1.

  For some lines of Ca\,I and Ca\,II we replaced original values of $\log\,(gf)$ 
and van der Waals damping parameters by more accurate values -- see 1,
4 and 8 columns of Tabl.\,4 of Mashonkina et al.  (2007) paper.  The spectra
of the hot spot and the photosphere were simulated with a spectral
resolution $R=600\,000$ and for a microturbulence velocity $V_{mic}=2$
km\,s$^{-1}$.  To align the spectral resolution with the observed one, the
theoretical spectra were broadened by a convolution with a gaussian, a
halfwidth of which was adjusted for each order of the echelle spectrum.

  To test our code we compared intensity spectra, simulated by the initial
program and by our code, in which we set $b_i=1$ for all lines.  The results
completely coincided.  We have also compared our calculations of three Ca\,I
lines profiles for the model of the solar atmosphere with the profiles,
simulated by L.\,I.  Mashonkina via the SIU program (Mashonkina et al.,
2007).  The difference of about $\sim 2\%$ in line's depth was found in the
case of Ca\,I 6439\,{AA} line, for which the difference between LTE and
non-LTE profiles was 20\%.  The depths for the rest considered lines differ
even less.  Aforesaid allows us to conclude that the modified SYNTHE code
works properly.

  The additional argument confirming that the program of calculating hot
spot spectra works properly looks as follows.  Typical energy of photons in
optical band is $\sim 2$ eV, and the temperature in the formation region of
Ca lines is $\sim 0.7$ eV (see Fig.,\ref{regionNCLF}), what means that
$\exp(h\nu/kT) \gg 1.$ Then:
$$
I_{nLTE} \sim
S_{nLTE}(\tau_{nLTE}=1) \sim
{\left( {b_u\over b_l} \right)}_{nLTE} \,B_\nu \left[T(\tau_{LTE}=1)\right]
\sim {\left( {b_u\over b_l} \right)}_{nLTE} \, I_{LTE}.
$$
We calculated $I_{nLTE}/I_{LTE}$ values for several lines of Ca\,I and
Ca\,II from this equation using $b$-factors calculated by our DETAIL code. 
It appeared that these quantities differed from the same values, derived from
our version of the SYNTHE code, less than 20\%.

  Let discuss now how departure from LTE affects on the intensity spectrum of
Ca lines, which are formed in the hot spot, considering it as a
plane-parallel slab.  If to fix parameters of undisterbed atmosphere then
spot's radiation intensity depends only on the parameters of the
accretion shock $V_0,$ $N_0$ and cosine $\mu$ of angle between the line of
sight and the normal to the slab.

%
\begin{figure}[h!]
  \begin{center}
  \includegraphics[scale=0.5]{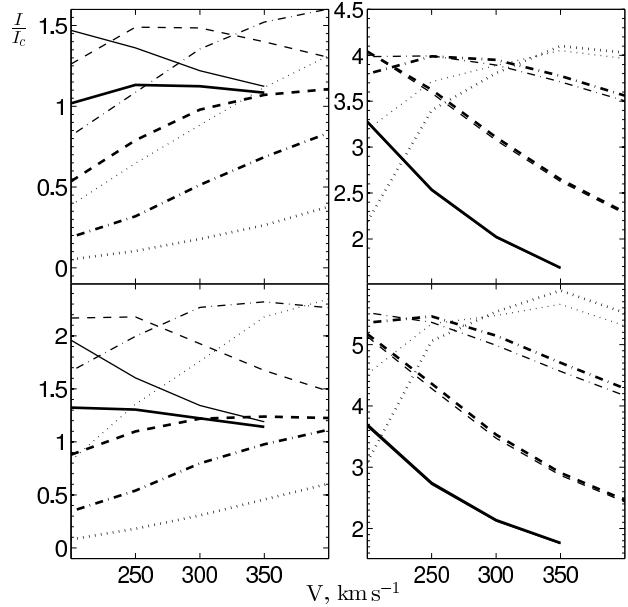}\\
 \caption{The intensity of Ca\,I 5589 (left column) and Ca\,II 8498 (right
column) lines at the central wavelength, normalized to the continuum
intensity, as a function of a pre-shock velocity of the gas $V_0$
km\,s$^{-1}$ for various values of $\log N_0:$ the solid, dashed,
dash-dotted, and dotted curves are for $\log N_0 = 13.0,$ $12.5,$ $12.0,$
and $11.5$ respectively.  The upper and lower rows correspond to $\mu=1$ and
$\mu=0.5.$ The thicker and thinner lines correspond to non-LTE and LTE
calculations respectively.  The parameters of the undisturbed atmosphere
are: $T_{ef}=4500$~X, $\log g=4.0.$ See text for details. 
} 
 \label{picint}
  \end{center}
\end{figure}
%

  To characterize a  behavior of a line, we choose a parameter $\xi,$ which
is defined as a ratio of the line intensity at the central wavelength to a
continuum intensity at $\lambda=\lambda_0:$ if $\xi<1$, then the line is an
absorption one, and if $\xi>1$, then it is an emission line.  $\xi$ values
for Ca\,I 5589 and Ca\,II 8498 lines are plotted on Fig.\,\ref{picint} as a
function of the accretion parameters $V_0,$ $N_0$ and the cosine $\mu.$ For
instance, it follows from the figure that the line Ca\,I 5589 can appear in
an absorption as well as in an emission, while the line Ca\,II 8498 is always
an emission line.

  We also plotted on the same figure the parameter $\xi,$ calculated in LTE
approximation: it can be seen that the differences between LTE and non-LTE
values of $\xi$ decreases, when $N_0$ and $V_0$ increase.  It looks strange
at the first glance because, according to (\ref{Fac}), with increasing of
$N_0$ and $V_0$ the accretion flux $F_{ac}$ also increases, and therefore, the
role of radiative processes, which produce the departures from LTE, should be
larger. But at the same time the gas density (and therefore, collisional rates)
in the hot spot also increases, because the pressure at spot's upper boundary
is proportional to $N_0 V_0^2$ (Dodin and Lamzin, 2012).

  We have seen on the example of Ca\,I 5589 and Ca\,II 8498 lines that the
parameter $\xi$ is a function of $V_0$ as well as $N_0$, however it turned
out that the functions $\xi=\xi(V_0,N_0)$ were very similar for all optical
Ca lines at considered parameters of the accretion flow.  Hence nearly
identical changes in the intensities of a pair of Ca lines can be obtained
by variation of $N_0$ as well as of $V_0.$ To a larger extent this is
correct for pairs of subordinate lines of Ca\,I, to a lesser extent -- for
pairs of Ca\,I-Ca\,II or pairs, one of which is Ca\,I 4227 resonant line. 
At practice it does not allow to determine separately $N_0$ and $V_0$
parameters from hot spot's spectrum if to use Ca lines only.


\section{The calculation of He\,I and He\,II spectrum}

  Having computed the level populations of helium, we calculated an emerging
spectrum as follows.  It was assumed that all background radiation with an
intensity $I_{\nu}^0$ at the frequency of considered spectral line was
formed in the layers below this line's formation region.  Then the solution
of radiative transfer equation for a slab with a thickness $H$ in the
direction, for which a cosine of an angle to the normal to the slab equals
to $\mu>0,$ can be written as:
$$
I_{\nu}(\mu) =
I_{\nu}^0(\mu) e^{-\frac{\tau_{\nu}(H)}{\mu}} + \frac{1}{\mu}
\int\limits_0^H{j_{\nu}(z)e^{-\frac{\tau_{\nu}(z)}{\mu}}{\rm d}z},
$$
and the total emission coefficient:
$$
j_{\nu}(z) =
\frac{h{\nu}N(z)\xi_{He}n_{U}(z)}{4\pi g_U}
\sum\limits_i A_{u}^{i}g_u^{i} \Psi \left[ z, \nu-\nu_i\left\{1-\frac{V(z)}{c}\mu\right\} \right].
$$

 The following notation were used in these expressions. $\nu_i$ is a
frequency of $i$-th component of a fine structure, $A_{u}^i$, $g_u^i$ are
Einstein A coefficient and statistical weight of an upper sublevel of the
$i$th component, and $g_U$, $n_{U}$ are statistical weight and relative
population of the upper level as a whole.  Corresponding atomic data were
adopted from the CHIANTI v.5.0 atimic database (Dere et al., 1997; Landi et
al, 2006).  Note by the way that line's central wavelength was defined as
weighted average over wavelengths of all $i$th components of the fine
structure with $(gf)_i$ weights.  $N(z)$ and $V(z)$ are a number density and
a velocity of the gas, which were adopted from models of post-shock cooling
zone (Lamzin, 1998) and hot spot (Dodin and Lamzin, 2012).  The velocity of
the gas settling $V(z)$ was equated to zero in the hot spot.  Helium
abundance $\xi_{He}=0.1.$

   $\Psi(z,\nu)$ is line's absorption and emission profile normalized as
$\int\Psi(z,\nu){\rm d}\nu=1.$ For He\,I lines we took into account Stark
broadening according to Dimitrijevic \& Sahal-Brechot (1984).  In the case
of He\,II lines we used Doppler profile at $T>3\times10^5$ K and at lower
temperatures (it corresponds to higher densities in our models) we took in
to account Stark broadening by an interpolation of Sch{\"o}ning and Butler
(1989) tables.  The tables cover the whole necessary density range, but
there is no data for $T<10^4$~K.  In this case we formally used the same
profile as for $T=10^4$ K, but it cannot lead to significant error, because
He\,II is practically absent at $T<10^4$ K -- see Fig.\,\ref{structure}.

The absorption coefficient was calculated as follows:
$$
\kappa_{\nu}(z) = \frac{\pi e^2}{m_ec}N(z)\xi_{He}\left[\frac{n_{l}(z)}{g_L} -
\frac{n_{u}(z)}{g_U}\right] \sum\limits_i(gf)_i \Psi
\left[ z, \nu-\nu_i\left\{1-\frac{V(z)}{c}\mu\right\} \right].
$$
where $n_{l}$, $n_{u}$, $g_L$, $g_U$ are the populations and statistical
weights of lower and upper levels, which are not splitted into components of
the fine structure.  The optical depth of a line in this case differ a
little from the same quantity, derived in the program of the level population,
because we did not consider there line's fine structure and assumed that the
line has Voigt profile.

   The method of a calculation of $I_{\nu}^0$ was described in the previous
section.  The calculations of $I_{\nu}$ were carried out for the same set of
$\mu$-values and on the same frequency grid as for $I_{\nu}^0.$

  We found from the comparison of He\,I and He\,II lines fluxes, calculated
with our code and with the Cloudy08 code, that all differences in the
results were completely caused by differences in the level populations due to
differences in used atomic data (see above).  This allows us to conclude
that our code works properly.

  We will use in what follows the model of a round homogeneous accretion
spot, i.e.  will assume that the accretion flow has the shape of a circular
cylinder, across which $N_0$ and $V_0$ parameters are the same everywhere. 
At first we separately calculated spectra of the hot spot and undisturbed
atmosphere.  Then we choosed relative area $f$ of the spot and its position
at stellar disk, defined by the angle $\alpha$ between spot's axis of
simmetry and the line of sight, and calculated resulting spectrum of the
"star $+$ spot"{} system by integration of specific intensity over visible
stellar hemisphere -- see Dodin \& Lamzin (2012) for details.

  We will illustrate the results of our simulation of helium spectrum on the
example of He\,II 4686 line.  Note by the way that CTTS are late-type stars
so helium lines are absent in their photospheric spectra.  Consider a set of
spectra of non-rotating star with $T_{ef}=4000$~K, $\log g =4.0,$ which has
an accretion spot on its surface.  Free parameters of the problem are
filling factor $(f=0.01,$ 0.03, 0.06, 0.1, 0.15, 0.2), spot's position
$(\alpha=0^o,$ $60^o)$ and parameters of the accretion flow: $\log N_0$ from
11.5 to 13.0 with a step of 0.5, $V_0$ from 200 to 400 km\,s$^{-1}$ with a
step of 50 km\,s$^{-1}.$

  He\,II 4686 line will be characterized by an equivalent width of a part of
its profile $\Delta v$ from -30 to 30 km\,s$^{-1}$, taking into account all
photospheric lines inside this velocity range.  The accretion power will be
characterized by the veiling $r_c$ of a photospheric spectrum due to a
continuum emission only rather than total veiling $r$, because it varies
from line to line.  Let us remind that $r_c$ is defined as the ratio of the
flux from accreting star $F_\lambda(\lambda)$ to that of non-accreting star
$F_\lambda^0(\lambda)$:
$$
r_c (\lambda) = \frac{F_\lambda}{F^0_\lambda}.
$$
The relation between $r_c$ and total veiling $r,$ defined by
Eq.\ref{er-definit}, depends on the accretion flow parameters: at small
values of $F_{ac}$ the veiling is caused by lines, and $r$ is several times
greater than $r_c$ (Dodin and Lamzin, 2012).

\begin{figure}[h!]
 \begin{center}
  \includegraphics[scale=0.65]{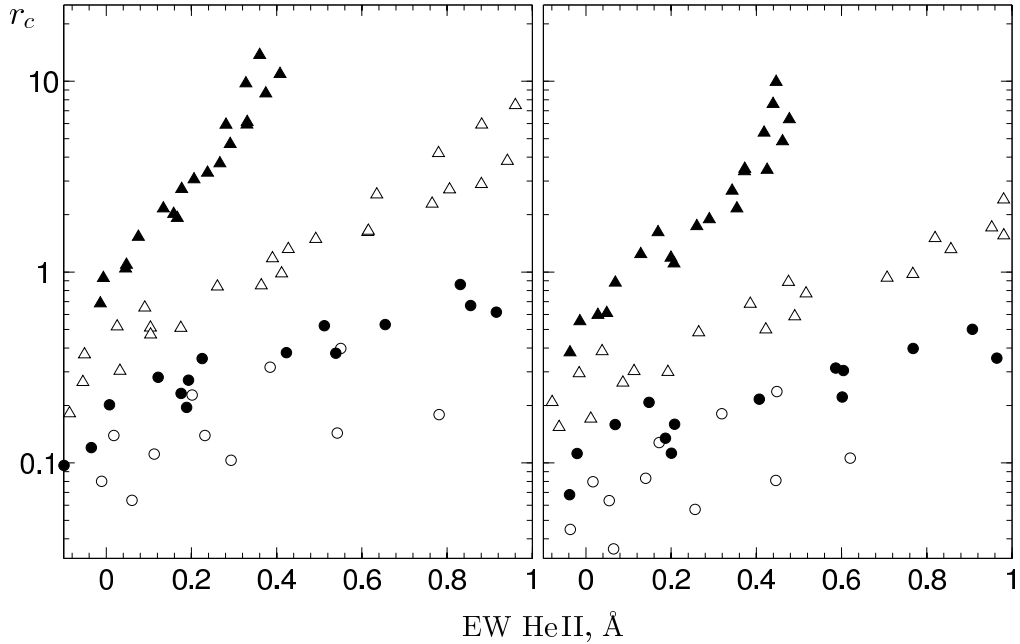}\\
  \caption{The veiling $r_c$, caused by a continuum emission, in the
vicinity of He\,II 4686 line as a function of its equivalent width.  The
models were simulated for the values of $V_0$ from 200 to 400 km\,s$^{-1}$
with a step of 50 km\,s$^{-1}$ and $f=0.01,$ 0.03, 0.06, 0.1, 0.15, 0.2 at
values of $\log N_0=11.5$ (empty rounds), $12.0$ (filled-in rounds), $12.5$
(empty triangles), $13.0$ (filled triangles).  The left and right panels of
the figure for the cases where the line of sight makes, respectively, the
angles $\alpha=0^o$ and $\alpha=60^o$ with the axis of the spot. Parameters
of a star are $T_{ef} = 4000$~K, $\log g = 4.0$.  
} 
 \label{he_rc}
 \end{center}
\end{figure}

   It appeared that at a fixed angle $\alpha$ the relation between $r_c$ and
He\,II 4686 line equivalent width is sensitive to infall gas density $N_0.$
It can be seen from Fig.\,\ref{he_rc} that models with the same value of
$N_0,$ but with various other parameters $V_0,$ $f$, form bands, which are
well separated at the chosen $\lg N_0$ step of 0.5, especially at a high
veiling.

%
%

The veiling $r_c$ is a purely theoretical parameter, because only the upper
limit of $r_c$ can be found from observed spectrum as a minimal $r$ of all
photospheric lines in the spectral region of interest.  Therefore, the
Fig.\,\ref{he_rc} is no more than an illustration of a sensitivity of some
spectral details to parameters of the accretion flow, in our case, to gas
density $N_0.$


\section{A comparison of the simulated spectra with observations}

  High resolution spectra of CTTS, which we compared with calculations, were
adopted from the Keck Observatory Archive 
http://www2.keck.hawaii.edu/koa/public/koa.php and (in the case of TW Hya)
from the VLT archive http://archive.eso.org/eso/eso\_archive\_main.html.  We
used automatically extracted and calibrated spectra, because did not found
any artifacts, which can influence to our conclusions.  Each observed
spectrum was transformed into the stellar rest frame by shifting it as a
whole (by radial velocity) until a coincidence of photospheric lines
positions with that of theoretical spectra.  In order to compare the
observed and simulated spectra of CTTS, we normalized both to continuum
level.  It appeared difficult to derive continuum level at wavelengths
shortward 4000 {\AA} for all stars except TW Hya. Due to this reason we
compare spectra of the stars with our simulations only at wavelengths
$\lambda>4000$ {\AA}, what, in the particular, did not allow us to use
Ca\,II H and K lines for a diagnostic.

  Even in the model of a homogeneous circular spot there are quite a lot
parameters that characterize a star ($T_{ef}$, $\log g$, $V_{mic},$
an equatorial rotational velocity $V_{eq}$, an inclination $i),$ an
accretion shock $(V_0,$ $N_0$) and a spot ($f,$ $\alpha).$ At
first we would like to describe, how we determined these parameters for each
spectrum of CTTS.

  The stellar parameters were adopted from the literature, except a few cases,
when we derived them by ourself to reach better agreement with observations. 
Reliable determinations of the microturbulence velocity for CTTS are
apparently absent and we always assumed $V_{mic}=2$ km\,s$^{-1}$.  The
parameters of 9 stars, spectra of which have been compared with calculations
are collected together in the Table \ref{stars}.

%
\begin{table}[h!]
  \caption [] {Stellar parameters.}
 \label{stars}
\begin{tabular}{|c|c|c|c|c|}
\hline
  Star & $T_{ef},$ K & $\log g$ & $i,^o$ & $V_{eq},$ km\,s$^{-1}$\\
\hline
  GM Aur & $4500^b$ & $4.0^k$ & $50^j$ & $17^h$  \\
  BP Tau & $4000^a$ & $3.5^a$ & $50^a$ & $15^a$  \\
  DK Tau & $4250^b$ & $4.0^b$ & $50^g$ & $15^h$  \\
  DN Tau & $4250^b$ & $4.0^b$ & $30^e$ & $17^f$  \\
  GI Tau & $4250^i$ & $4.0^g$ & $70^g$ & $9^g $  \\
  GK Tau & $4250^k$ & $3.5^g$ & $50^g$ & $20^g$  \\
V836 Tau & $4500^k$ & $4.0^k$ & $60^k$ & $16^k$  \\
  DI Cep & $5500^k$ & $3.5^c$ & $60^c$ & $25^c$  \\
  TW Hya & $4000^d$ & $4.0^d$ & $10^d$ & $16^d$  \\
\hline
\multicolumn{5}{p{15cm}}{ \footnotesize
References:
a -- Donati et al. (2008);
b -- Schiavon et al. (1995);
c -- Gameiro et al. (2006);
d -- Donati et al. (2011);
e -- Bouvier and Bertout  (1989);
f -- Smith (1994);
g -- Johns-Krull and Valenti (2001);
h -- Hartmann and Staufer (1989);
i -- Kenyon and Hartmann (1995);
j -- Gr{\"a}fe et al. (2011);
k -- $T_{ef},$ $\log g,$ $V_{eq}$ were chosen by us, and $i$ adopted arbitrary.
}\\
\end{tabular}
\end{table}
%

  The parameters of an accretion shock and a hot spot were found as follows. 
Pre-shock gas number density $N_0$ was determined from the comparison of
veiling $r_c$ in the vicinity of He\,II 4686 line and $EW$ of this line (see
the previous section and Fig.\,\ref{he_rc}).  One can assert that an error
of $N_0$ is significantly less than the density step of our shock models
grid, which is equal to 0.5 dex, because it appeared impossible to
reproduce simultaneously observed veiling and $EW$ of He\,II 4686 line for
such deviations from derived value of $N_0$ by variation of other
parameters.

\begin{figure}
 \begin{center}
  \includegraphics[scale=0.65]{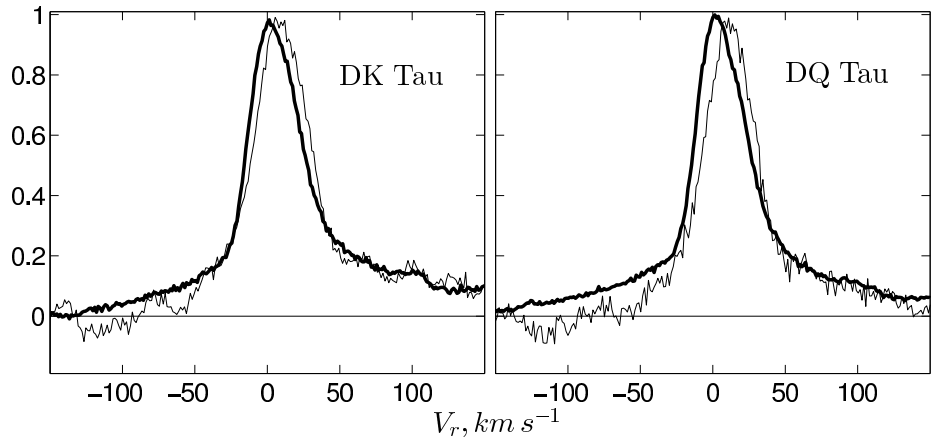}\\ 
   \caption{ Profiles of He\,I 5876 (thick curve) and He\,II 4686 (thin
curve) lines normalized to their maximum in spectra of DK Tau (JD
2\,454\,487.872) and DQ Tau (JD 2\,455\,144.123) after the substraction of
the veiled photospheric spectrum. }
 \label{DKDQ}
 \end{center}
\end{figure}

  We discovered that He\,II 4686 line consist of two components: one is strong
and narrow, and second is broad and weak.  To our knowledge, the broad
component of He\,II 4686 line has never been described before, possibly
because it is masked by a blend of photospheric lines. But if to
subtract calculated veiled photospheric spectrum from the observed one, then
it turns out that the profile of He\,II 4686 line is practically coincide
with that of He\,I 5876 line (see Fig.\,\ref{DKDQ}), for which an existence
of the broad component is well recognized.  We used only a central peak and a
blue wing of He\,II 4686 line to derive $N_0.$

%
%

  Thus, it can be concluded that broad components of both He lines in
considered stars are formed in the same region.  As far as broad components
of the lines are redshifted, it is naturally to assume that they are formed
somewhere in pre-shock accretion flow, which links the disc and the star.

\begin{figure}
 \begin{center}
  \includegraphics[scale=0.6]{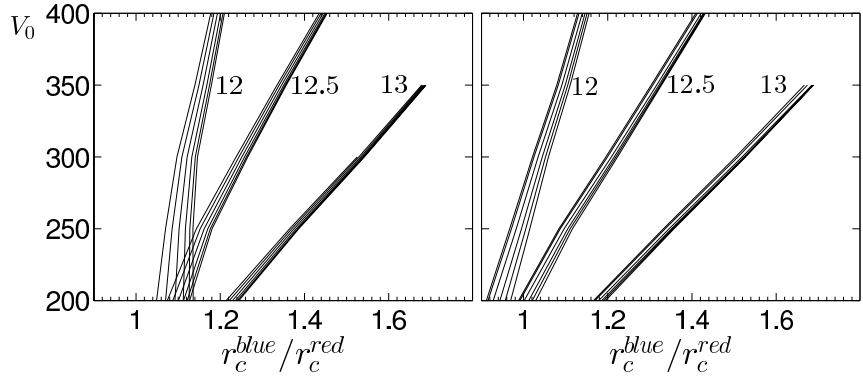}\\ 
   \caption{Dependence of pre-shock velocity $V_0$ (in km s$^{-1}$) on
the ratio of veilings $r_c$ in the blue (near the He\,II 4686 line) and
red (6000-6500 {\AA\AA}) spectral bands.  Different groups of lines
correspond to various values of $\log N_0$ (written near the group), and
various lines within each group correspond to various filling factors. 
Models at the left panel corresponds to spot's orientation $\alpha=0^o$ and
at the right panel -- to $\alpha=60^o.$ The parameters of the star are:
$T_{ef}=4000$ K, $\log g = 4.0.$ }
 \label{vel}
 \end{center}
\end{figure}

  Having derived $N_0,$ it is possible to determine $V_0$ by analizing 
SED of the veiling continuum, which can be characterized by
$r_c=r_c(\lambda).$ It is caused by the fact that hot spot effective
temperature and $r_c(\lambda)$ depend on $V_0$ only if $N_0$ is fixed. 
Aforesaid is illustrated by Fig.\,\ref{vel}, on which $V_0$ is plotted as a
function of a ratio of the veilings $r_c$ in the vicinity of He\,II 4686
line and in the region 6000-6500 {\AA\AA} for the model of a star with
$T_{ef}=4000$ K, $\log g = 4.0$ and the same parameters of the accretion
spot, as on Fig.\,\ref{he_rc}.  It is seen that for the chosen spectral band
$V_0$ can be derived more precisely in the case of high density and the
result weakly depends on $\alpha$ and $f.$ The figure is presented as an
illustration of the method only: in reality we fitted not only
$r_c(\lambda)$ but also total veiling $r=r(\lambda)$ in the whole observed
spectrum.

%
%


  Note that while hot spot models were calculated at a discrete grid of the
parameters $V_0$ (200-400 km\,s$^{-1}$ with a step of $50$~km\,s$^{-1})$ 
and $\log N_0$ (11.5-13 with a step of 0.5 dex), we calculated spectra also for
intermediate values of $V_0,$ $N_0,$ obtained by means of 2-D linear
interpolation of specific intensity $I_\nu(\mu,\lambda).$

\begin{figure}[h!]
 \begin{center}
  \includegraphics[scale=0.5]{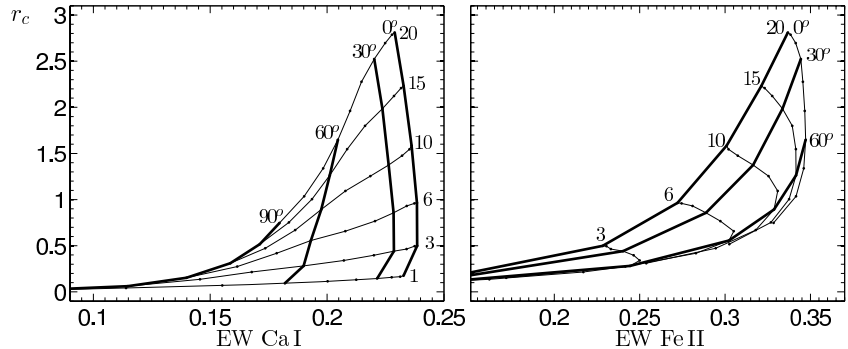}\\
  \caption{ $r_c$ value in the vicinity of Ca\,I 4226.73 
(left panel) and Fe\,II 4233.17 (the right panel) lines as function of EW
(in \AA) of these lines (see text for details).  Thin curves correspond to
models with the same relative area $f,$ values of which (in \%) are
indicated at the figure. Dots at these curves mark models with $\alpha$
that differs in $10^o.$ Bold curves connect models with different $f$ but equal
$\alpha$. Additional parameters of the models are: $T_{ef}=4000$~K, $\log
g=4.0,$ $\log N_0=13.0,$ $V_0=200$ km\,s$^{-1}$.  }
 \label{angle}
 \end{center}
\end{figure}

  Having derived accretion shock parameters $N_0$ and $V_0,$ one can find
parameters of the spot.  An observed quantity, suitable for this purpose, is
veiling by lines or, in other words, an intensity of the narrow component of metal's
emission lines, which fill in to various extent respective photospheric
lines.  Different brightening low $I_\nu(\mu)$ of lines and continuum allows
to determine the angle $\alpha$ and filling factor $f$ from the relations
between equivalent widths of narrow components of emission lines and $r_c$ value.
Fig.\,\ref{angle} illustrates this statement on an example of two lines: 
Ca\,I 4226.73 resonant line, which always has an emission core in CTTS
spectra and Fe\,II at 4233.17 {\AA} line located near the first.

%
%

  The narrow emission components of both lines are located inside the broad
(photospheric) Ca\,I 4226.73 line, therefore it is better to measure their
equivalent widths relative to the flux at the nearby point of the
photospheric line wing, rather than relative to uncertain continuum level. 
These points were chosen in theoretical spectra at $\lambda = 4226.63$ {\AA}
and $\lambda = 4233.30$ {\AA} for Ca\,I 4226.73 and Fe\,II 4233.17 lines
respectively.

  When suitable values of $\alpha$ and $f$ were choosen to reproduce
observed profile and intensity of Ca\,I 4226.73 line the model with solar
calcium abundance in accreted gas $\xi_{Ca}$ also reproduced these
quantities for all other Ca\,I lines as well as for Fe\,II 4233.17 line. 
The only exception was TW Hya for which we tried to use models with 
$\xi_{Ca}$ 3 and 10 times less than solar value.

  The using of $\xi_{Ca}$ as a free parameter and LTE-spectra of Fe\,II in
our approach makes determination of $f$ and $\alpha$ values somewhat
uncertain, however there is no sense to determine these parameters with high
accuracy in the frame of simple round homogeneous spot model.  Due to the
same reason we choose the best fit model by means of visual comparison of
simulated spectra with observed, i.e.  did not use any mathematical
optimization technique.

  However it should be noted that the model allows in principle to
determine not only the angle $\alpha,$ but a latitude $\theta$ and a
longitude $\varphi$ of spot's center on the stellar surface, using
the relation:
$$
\cos \alpha = \cos \theta \cos i + \sin \theta \sin i \cos \varphi,
$$
where $i$ is an inclination of the rotational axis of the star to the line
of sight.  We measured the angle $\theta$ from the nearest pole, and the
angle $\varphi$ from the central meridian in the direction of stellar
rotation.

   Spot's position at stellar surface relative to the Earth charcterized by
$\alpha$-angle varies periodically due to variations of $\varphi,$ while $i$
and $\theta$ angles do not vary.  It means formally that parameters $i,$
$\theta$ and $\varphi$ can be found by an analysis a few $(\geqslant 3)$
spectra, obtained at different rotational phases.

\begin{figure}[h!]
 \begin{center}
  \includegraphics[scale=0.5]{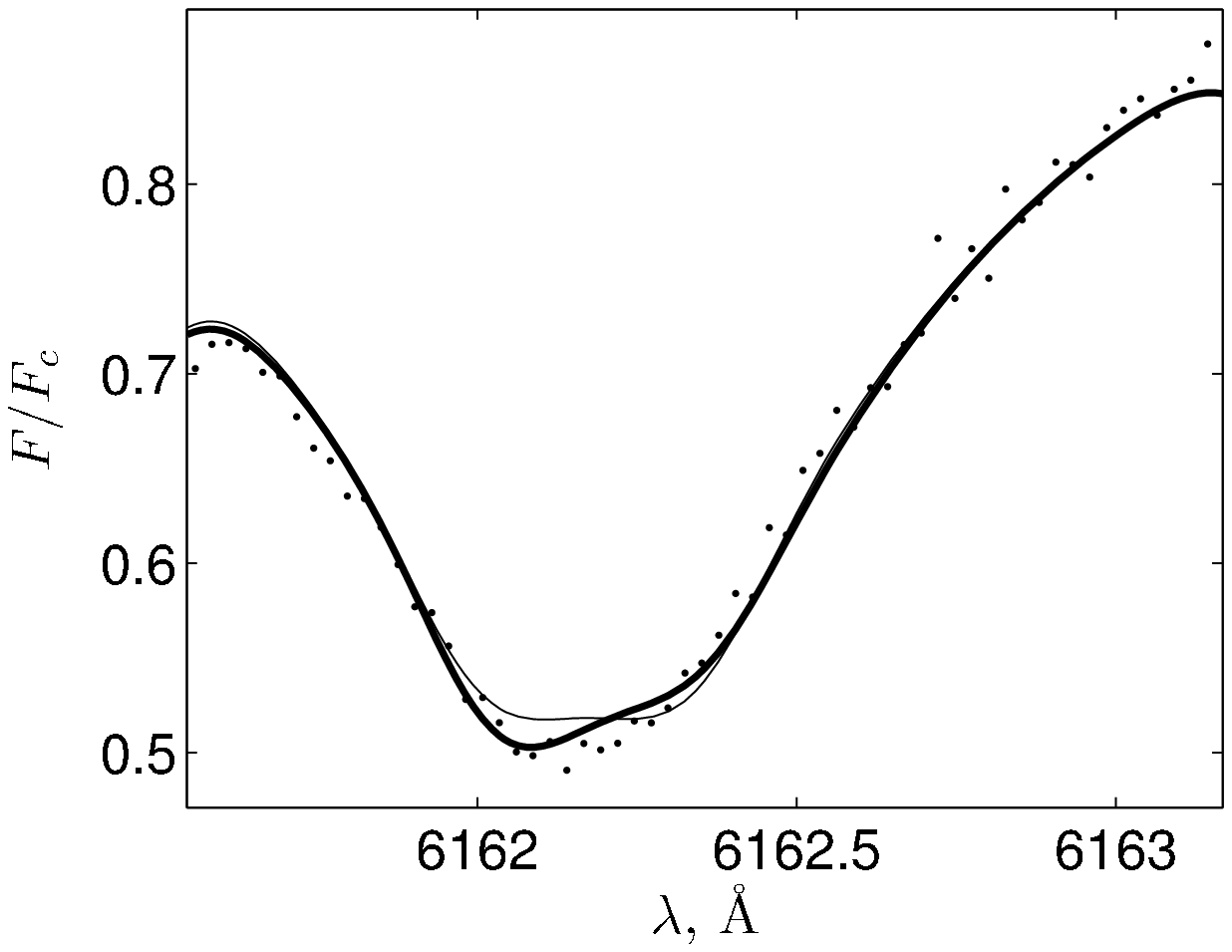}\\
 \caption{ Comparison of Ca\,I 6162.17 line profile in DK Tau spectrum at JD
2\,454\,487.866 (dots) with simulated profiles: the thick and thin curves
correspond to models with $\varphi=15^o$ and $\varphi=0^o$ respectively. 
Other parameters of the models can be found in Tabl.\,\ref{stars} and
Tabl.\,\ref{models} } 
\label{phi}
 \end{center}
\end{figure}

  In addition in spectra of rotating star the position of an emission component
inside respective absorption line varies with time due to Doppler effect. 
Therefore, observed line profiles should looks different at different
$\varphi$.  The described effect is manifested in the form of periodical
variations of radial velocities of CTTS photospheric lines (Zaitseva et al.,
1990; Petrov et al.,2001). Fig.\,\ref{phi} shows that the shape of
absorption lines in high-quality spectra can be sensitive enough to the
angle $\varphi.$ If an asymmetry of photospheric lines profiles was
noticeable in the observed spectrum, we determined $\theta$ and
$\varphi,$ instead of $\alpha$ (for illustrative purposes only) 
adopting $i$ from the literature -- see table \ref{stars}.
 
%
%

  But in some analyzed spectra profiles of photospheric lines had a
symmetrical shape, while the veiling by lines was noticeable, judging from
the theoretical spectrum.  Apparently it means that accretion zone (spot) of
these stars is extended in longitude in agreement with numerical simulations
(Romanova et al. 2004).  We posed $\varphi=0$ in such cases and determined 
$\theta$ only.

%
\begin{table}
  \caption{Acretion spot parameters}
 \label{models}
\begin{center}
\begin{tabular}{c|c|c|c|c|c|c|c|c}
\hline
The star & JD 245... & $V_{0}$ & $\log N_0$ & $f$, \% & $\theta$,$^o$ &
$\varphi$,$^o$ & $L_{ac}/L_*,$  \% & $\dot M_{ac}$ \\
\hline
GM Aur & 3337.850  & 280 & 13.0 & 1.5 & 0  & 0   & 15 & 8.8 \\
       & 3984.035  & 290 & 13.0 & 1.6 & 0  & 0   & 18 & 9.7 \\
       & 4717.048  & 300 & 13.0 & 1.4 & 0  & 0   & 18 & 8.8 \\
\hline
BP Tau & 1883.003  & 300 & 13.0 & 2.0 & 30 & 0   & 40 & 13 \\
       & 2535.982  & 350 & 13.0 & 2.0 & 20 & 0   & 64 & 15 \\
       & 3301.985  & 325 & 12.9 & 1.3 & 0  & 0   & 27 & 7.2 \\
       & 3368.772  & 350 & 13.0 & 0.8 & 30 & 0   & 26 & 5.9 \\
       & 3984.022  & 300 & 12.6 & 3.0 & 20 & 0   & 24 & 7.5 \\
       & 4487.837  & 350 & 13.0 & 1.3 & 20 & 0   & 42 & 9.6 \\
       & 4544.739  & 335 & 13.0 & 2.0 & 20 & 0   & 56 & 14 \\
       & 4717.052  & 350 & 13.0 & 1.2 & 20 & 0   & 39 & 8.8 \\
       & 4807.915  & 350 & 13.0 & 1.5 & 20 & 0   & 48 & 11 \\
\hline
DK Tau & 4487.866  & 300 & 11.9 & 12  & 40 & 15  & 15 & 6.0 \\
       & 4487.872  & 300 & 11.9 & 12  & 40 & 20  & 15 & 6.0 \\
\hline
DN Tau & 5135.041  & 230 & 12.2 & 3.0 & 30 & 60  & 3.4 & 2.3 \\
       & 5284.753  & 280 & 13.0 & 1.2 & 20 & 50  & 16 & 7.1 \\
\hline
GI Tau & 4487.930  & 330 & 13.0 & 2.0 & 40 & -40 & 42 & 14 \\
\hline
GK Tau & 4487.924  & 270 & 13.0 & 1.5 & 10 & 0   & 17 & 8.5 \\
\hline
V836 Tau & 4806.967& 230 & 12.5 & 5.0 & 30 & 60  & 9.0 & 7.6 \\
\hline
DI Cep & 4609.131  & 260 & 13.0 &  4  & 40 & -10 & 15 & 22 \\
\hline
TW Hya & 4157.875  & 350 & 13.0 & 1.0 & 10 & 0   & 32 & 7.3 \\
       & 4158.607  & 360 & 12.8 & 1.0 & 10 & 0   & 22 & 4.8 \\
\hline
\multicolumn{9}{p{13cm}}{\footnotesize Note. --
$V_0$ in km\,s$^{-1}$, $N_0$ in cm$^{-3},$
$\dot M_{ac}$ in ${(R_*/R_{\odot})}^2 \times 10^{-9}M_{\odot}/$year.
The quality of the spectra of BP Tau allows to reliably determine only the density $N_0.$
For TW Hya $\xi_{Ca}/\xi_{Ca}^\odot = 0.3$ and 1 for all other stars.
}\\
\end{tabular}
\end{center}
\end{table}
%

\begin{figure}[h!]
 \begin{center}
  \includegraphics[scale=0.5]{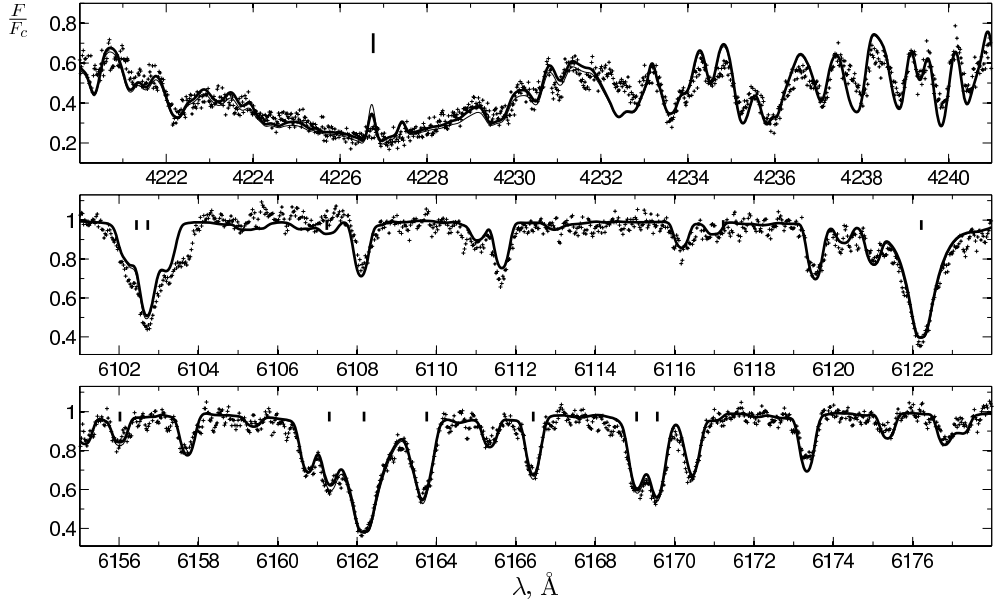}\\
 \caption{ Some regions of the spectrum of GM Aur at JD 2\,453\,337.850
(crosses), which contain Ca\,I lines, positions of which are marked by
vertical bars.  In the case the results of non-LTE (thick line) and LTE
(thin line) calculations at the same parameters of the accretion spot
practically coincide.  } 
\label{GM1}
 \end{center}
\end{figure}

   Following the described algorithm, we have determined accretion
parameters of all CTTS from the Tabl.\,\ref{stars} and presented them in the
Tabl.\,\ref{models}.  It can be seen from this table that the majority of
stars accretes a gas with a number density $N_0>3\times 10^{12}$ cm$^{-3}.$
It turned out that at such high densities LTE and non-LTE spectra were similar
and reproduced observed depth of all subordinate lines of Ca\,I in optical
band and the narrow emission component of Ca\,I 4226.73 resonant line.  An
example is the star GM Aur $(\log N_0 =13),$ for which parts of the spectrum
are shown on Fig.\,\ref{GM1}.  For every star except TW Hya, which will be
discussed below, an acceptable agreement between the calculations and the
observations was achieved at the solar abundance of calcium.  Moreover,
depths of photospheric lines of other elements are also reproduced in our models
with a good accuracy.

%
%

\begin{figure}[h!]
 \begin{center}
  \includegraphics[scale=0.5]{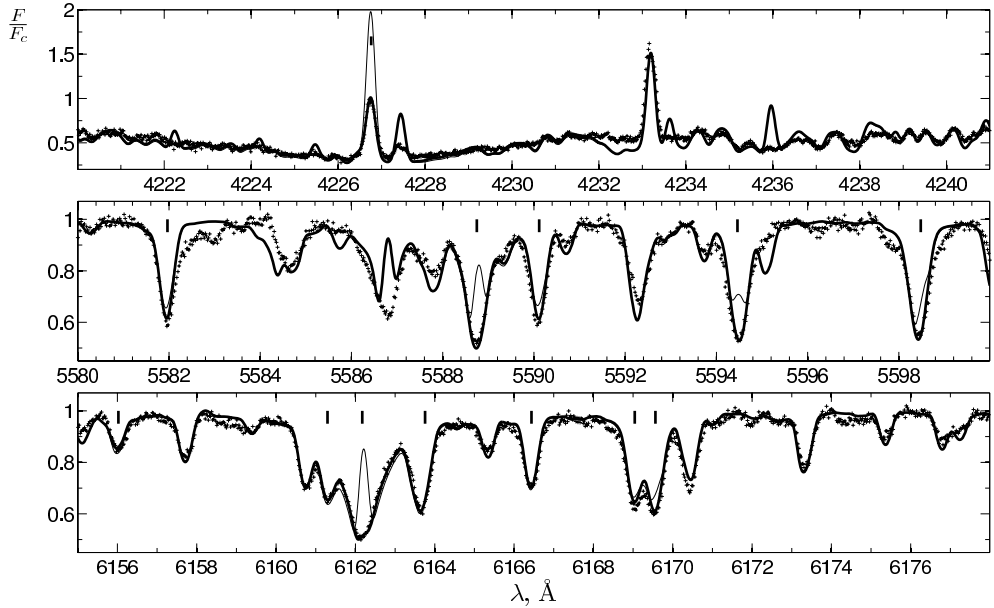}\\
  \caption{ The same as in Fig.\,\ref{GM1}, but for  DK Tau (JD
2\,454\,487.866).  The thick and thin curves correspond to a non-LTE
spectrum for calcium lines and a LTE spectrum for all lines at the same
parameters respectively.} 
\label{DK1}
 \end{center}
\end{figure}

   However, in the case of DK Tau, for which $\log N_0 = 11.9,$ the difference
between LTE and non-LTE models is very noticeable, see Fig.\,\ref{DK1}.  If
the model is chosen so that it reproduces the equivalent width of the
He\,II 4686 line and SED of the veiling continuum, then the narrow emission
components of Ca\,I at solar abundance of calcium are much stronger than the
observed ones.  The use of non-LTE calculations for calcium allows to
achieve a good agreement with observations without assumptions about the
depletion of this element.  The model are well reproduce a lot of absorption
lines of other elements, but for some lines of Fe, for instance Fe\,I 4235.9
and Fe\,I 5586.8, the model predicts a more strong emission component than
observed.  It indicates probably that it is necessary to consider non-LTE 
effects not only for calcium, but also for iron in the case of DK Tau.

%
%

\begin{figure}[h!]
 \begin{center}
  \includegraphics[scale=0.5]{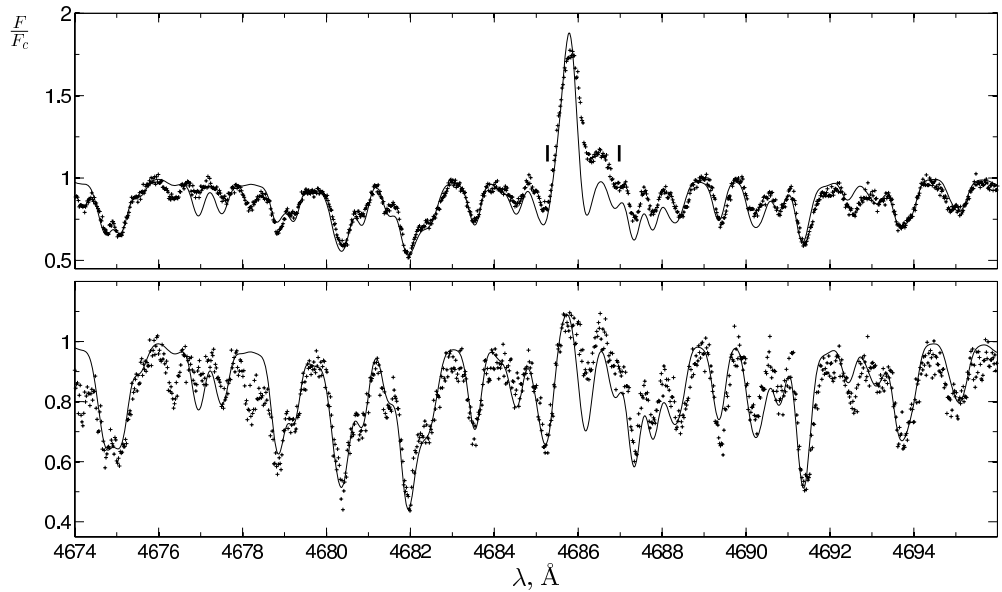}\\
  \caption{The region of the spectra of DK Tau (top panel) and GM Aur
(bottom panel) in a vicinity of He\,II 4686 line.  Julian dates and the
parameters of the models, the spectra of which are shown by the thin curve,
are the same as in Fig.\,\ref{GM1}, \ref{DK1}.  Positions of Ca\,I
lines are marked by vertical bars.} 
\label{HeII-spectra}
 \end{center}
\end{figure}

  Fig.\,\ref{HeII-spectra} shows how the models with the same parameters
reproduce observed spectra of GM Aur and DK Tau in the vicinity of He\,II
4686 line. The figure demonstrates also how gas density $N_0$ affects on EW 
of the line.

%
%

\begin{figure}
 \begin{center}
  \includegraphics[scale=0.5]{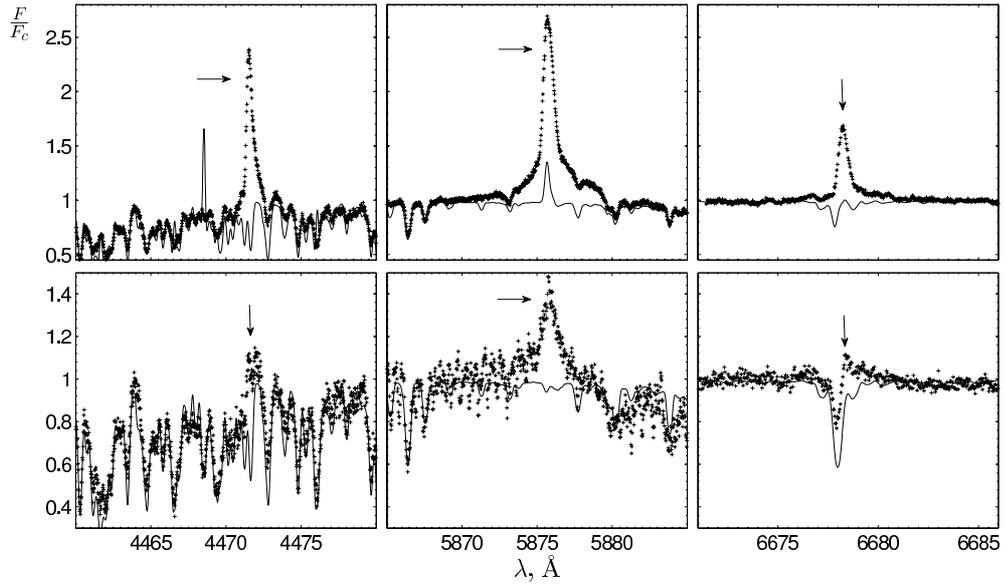}\\
  \caption{The region of the spectra of DK Tau (upper panel) and GM Aur
(lower panel) in the vicinity of He\,I 4471, 5876 and 6678 lines.  Julian
dates and parameters of the models, the spectra of which are shown by the
thin curve, are the same as for Fig.\,\ref{GM1} and \ref{DK1} respectively. 
The absorption feature at the left from He\,I 6678.16 line is photospheric FeI
6677.99 line.} 
\label{HeI-spectra}
 \end{center}
\end{figure}

  As we noted in the section, devoted to the calculations of helium level
population, a probable deviation of LTE-distribution of $T(\tau)$ from the
true structure in the upper layers of a hot spot and uncertanties in the
collisional cross sections from upper levels of He\,I do not allow to trust
to the simulated profiles of He\,I.  It turned out that intensities of He\,I
4471, 4713, 5016, 5876, 6678, 7281 {\AA\AA} lines predicted by our models
were significantly smaller than observed.  The ratios of an equivalent widths
of these lines are also disagree with observations especially in the case of
He\,I 4471 line ($n=4\rightarrow 2$ transition): its $EW$ in observed
spectra is comparable with $EW$ of 5876 {\AA} line $(n=3\rightarrow 2$
transition), while in the simulated spectra 4471 {\AA} line is significantly
weaker than 5876 {\AA} line and almost invisible in the spectrum.  Aforesaid
is illustrated by Fig.\,\ref{HeI-spectra} on examples of GM Aur and DK Tau.

%
%

   Unfortunately we could not use lines of Ca\,II for the diagnostic in the
case of CTTS, spectra of which were taken from Keck Observatory Archive.  In
spectra of these stars only Ca\,II H and K resonance lines and the infrared
triplet (IRT) at $\lambda=8498.0,$ 8542.1, 8662.1 {\AA}{\AA} have
significant intensity.  As we have noted before, the difficulties with
determination of continuum level at $\lambda < 4000$ {\AA} did not allow us
to use the H and K lines.  What about IR triplet lines they either were out
of observed spectral range or consisted almost fully from broad component so
that the narrow component cannot be extracted.  At the same time it is
necessary to note that other Ca\,II lines, which are absent in the spectra of
investigated stars, also have negligible intensity in spectra of our models.

  These problems have been overcome in the case of TW Hya, for which
spectra were retrieved from the VLT archive.  For this star the relative
intensity of the He\,II 4686 line and the SED of the veiling continuum were
successfully fitted by a model of the homogeneous spot, but having assumed
that the calcium abundance in the falling gas is three times smaller than
solar.  Note, that the age of TW Hya is about 8 Myr (Donati et
al., 2011), i.e. it is significantly older than other considered stars,
therefore, the depletion of calcium in the accreted  gas due to its
accumulation in heavy grains looks resonable.

%
\begin{figure}
 \begin{center}
  \includegraphics[scale=0.5]{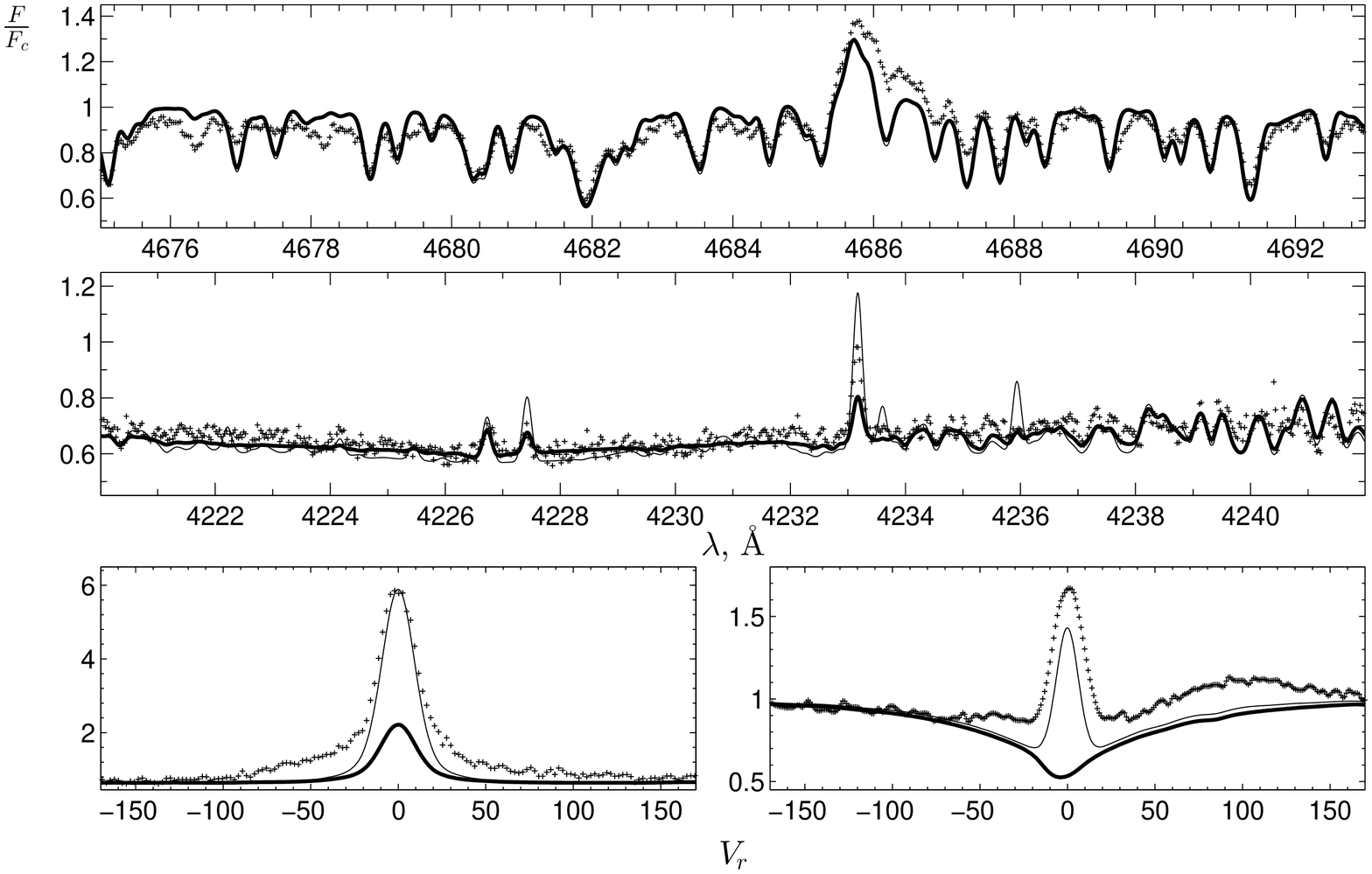}\\
  \caption{Some regions of TW Hya spectrum (JD 2\,454\,157.875)  in the
vicinity of He\,II 4686 line (upper panel) and Ca\,I lines (middle panel) as
well as observed profiles of Ca\,II K line $(\lambda = 3933.7$, left lower panel) and $\lambda =
8662.1$ {\AA} line of Ca\,II IR triplet (right lower panel). The crosses correspond to the observed specra. 
The thick curve corresponds to the model of the homogeneous spot with parameters from the
Tabl.\,\ref{models}.  The thin curve corresponds to the model of the
two-component spot, the parameters of which are given in the text.}
\label{twospot}
 \end{center}
\end{figure}
%

  However, as can be seen from the figure, the intensity of the narrow
component of the Ca\,II lines is significantly smaller than the observed.
In order to increase it, we need to decrease pre-shock density $N_0,$ 
but then the intensity of He\,II 4686 line became larger than the observed. 
This contradiction can be overcomed by assuming an inhomogeneity of the spot,
namely, by assuming that the spot with high densities and velocities of the
falling gas is surrounded by an accretion zone with lower values of $N_0$
and $V_0.$

  To illustrate it, we have calculated two-component spot's model for TW
Hya.  The model assumes that the central region with $V_0=370$ km\,s$^{-1}$,
$\log N=13.0,$ $f=0.008,$ $\varphi=0^o,$ $\theta=10^o$ is surrounded by a
coaxial ring zone with $V_0=250$ km\,s$^{-1}$, $\log N_0=11.5$ and $f=0.06.$
As in the case of the homogeneous spot, the abundance of calcium was adopted
3 times smaller than solar.  The simulated spectrum of two-component spot
are shown on Fig.\,\ref{twospot} by a thin curve.  We did not try to
choose the parameters of the composite spot in order to achieve the best
agreement with observations. The figure demonstrates only that 
inhomogenious spot model allows to reproduce the observed flux of the Ca\,II
lines, not disturbing the agreement between the calculations and the
observations for veiling and other lines.

 The ratio of accretion to stellar luminosities and mass accretion rate,
presented in two last columns of the Tabl.\,\ref{models}, were calculated
from the relations:
$$
{ L_{ac} \over L_*} = {f F_{ac} \over \sigma T_{ef}^4},
$$
$$
\dot M_{ac} = 4\pi R_*^2 f \mu m_p N_0 V_0,
$$
where $\mu\simeq 1.3$ is the average molecular weight, $m_p$ is the proton
mass.

   $\dot M_{ac}$ value in the table was calculated for stellar radius
$R_*=R_\odot,$ because we did not determine here $R_*$ and the values from
the literature is very unreliable.  Note that if there are two accretion
spots with similar parameters, which are located in diametrically opposite
regions on the surface of CTTS, then the values of $L_{ac}/L_*$ and $\dot
M_{ac}$ from the table should be increased by 2 times.

  Calvet and Gullbring (1998) investigated the first six stars from our
Tabl.\,\ref{models} in the frame of their model, which assumes that the spot
emits in continuum only. $F_{ac}$-values for these stars, obtained by
Calvet and Gullbring are, in average, twice larger than our, while filling
factors $f,$ vice versa, are a few times smaller. Note that, in contrast to
our paper, Calvet and Gullbring used $V_0$-values derived from the
assumption that they are free fall velocity at 5$R_*$ and adopting
masses and radii of stars found by Gullbring et al. (1998) from evolutionary
tracks. These $V_0$-values are systematically smaller than our.

  For five considered CTTS out of nine we have analyzed more than one
spectrum -- see Tabl.\,\ref{models}.  In the case of DK Tau and TW Hya the
spectra were taken with an interval less than one day, and the parameters of
the spot, derived from them, differ just a little.  A good repeatability
of the parameters can be noted also for GM Aur. The signal-to-noise
ratio of BP Tau spectra is relatively low, and due to this reason we can
reliably determine $N_0$ value only, which in one case out of nine
differs significantly from other.  Only for DN Tau the temporal variations
of the parameters were significant.  It can be assumed that in this case the
properties of the accretion stream were changed indeed, because these two
spectra are visually differs from each other.


\section*{Conclusion}

  It is generally recognised that emission spectrum of CTTS is a consequence
of a magnetospheric accretion of matter from a protoplanetary disk, but the
spectrum has not been reproduced quantitatively until now.  It raises doubts
about the reliability of the existing estimates of the parameters of the
accretion flow in CTTS, general characteristics of young stars and the
interstellar and/or circumstellar extinction $A_V$ for these objects.  A
simulation of CTTS spectrum is complicated by the fact that most of emission
lines consist of a narrow and a broad components, which are formed in
regions with very different physical conditions, geometry and velocity
field.

   The difference is so great that at a present day narrow and broad
components of the emission lines are being modelled independently of each
other.  It introduces an uncertainty in a process of a comparison of 
simulated intensity and profiles of one component due to the {\it a
priori} unknown contribution of the second.  This problem is especially
serious in the case of the broad components, because they are formed in a moving
gas of the magnetosphere of CTTS, therefore, separation of the components,
which is used now, by means of a decomposition of the observed profile into
a pair of gaussians cannot be considered as a serious basis for a comparison
of simulations with observations.  In addition, the simulation of profiles
of the broad component is a very laborious and multiparameter problem, which
requires a simultaneous solution of 3-D radiation transfer and MHD equations 
(Kurosawa and Romanova, 2012).

  We have seen on an example of He\,II 4686 and Ca\,II lines that the broad
component complicates comparison of the simulated spectrum of the hot spot
with observations.  At the same time, our calculations reproduce well
profiles of the most lines of neutral metals, for instance Ca\,I.  It means
that the broad component is almost absent in these lines.  If we reconstruct
the spectrum of the hot spot by using such lines, then the profiles of the
broad component can be found by substracting the simulated profiles from
observed ones.  In this paper such approach has allowed us for the first
time to discover the broad component in He\,II 4686 line.

  Nevertheless, the main purpose of our modeling is to determine a shape of
the accretion spot and distributions of the parameters $V_0$ and $N_0$
within it.  Moreover, the stellar parameters should be determined, namely:
$T_{ef},$ $R_*,$ $i,$ $V_{eq}$ and $A_V.$ This problem can be solved by
Doppler imageing, using the dependence of $I_\nu = I_\nu (\mu),$ calculated
for as wide as possible spectral range for a comprehensive grid of the
parameters.  Calculations, performed by Dodin and Lamzin (2012), and in this
work are the first steps in solution of this problem.

  A good agreement with observations allows to trust to the conclusion that
the most of considered stars accretes matter with pre-shock gas density
$>3\times 10^{12}$ cm$^{-3},$ although this conclusion we obtained in the
framework of the homogeneous circular spot model.  Our $N_0$ value found for
TW Hya coincides with an estimation of Kastner et al.  (2002) derived from
X-ray observations, but in the case of BP Tau Schmitt et al.  (2005) found
$N_0$ an order of magnitude smaller than our value.

  It turns out that at such density the spectrum of Ca\,I as well as of
other metals are close to LTE.  Taking into account a simplicity of our
model, it is difficult to say about a reliability of our conclusion about a
depletion of calcium in the falling gas in case of TW Hya.  However, TW Hya
is older than other considered stars in a few times, therefore the depletion
of calcium in the falling gas due to its accumulation in heavy grains is
possible exactly for this star.

  For further progress in the mapping of CTTS's accretion spots the
following should be done.

  1) The appropriate observational data should be prepared. At the beginning
all available spectra of CTTS should be carefully processed: recall that we
used spectra after automatic processing in this study.  The simulated
spectra have been compared with spectra, which were normalized to continuum
level for each echelle order, therefore the obtained results do not depend
on the interstellar reddening.  A good agreement of observed spectra with
simulated ones, for which $F_\lambda = F_\lambda(\lambda)$ dependence is
known, means that we know stellar spectrum, undistorted by interstellar
reddening.  Hence if we obtain from observations $F_\lambda(\lambda)$
dependence, then we could find not only $A_V,$ but also try to derive the
dependence $A_\lambda(\lambda).$ If the dependence $F_\lambda(\lambda)$ is
obtained in absolute units, then we could also find luminosity and radius of
a star.

  2) Non-LTE structure of upper layers of the hot spot should be calculated.
It will allow to calculate more accurately an emission spectrum of He\,I as
well as of hydrogen.  By the way, we calculated the non-LTE spectrum of 
hydrogen, but not used in the diagnostics due to two reasons:
at first, profiles of Balmer lines are dominated by the broad component and
secondly, the hydrogen lines are formed exactly in the upper layers of the hot
spot.

  Before this work we have belived that narrow component of H and He, in contrast
to Ca, are formed equally in the hot spot and in a moving gas in the
postshock cooling zone.  For an uniform calculation of the level populations
in regions with very different physical conditions we used the
escape probability approach.  However, it is clear now that the main flux
of these lines are formed in the hot spot, therefore in the future the
calculation of the level populations of He and H should be carried out just
like for Ca.

3) To identify as many as possible lines in CTTS spectra, which are
sensitive to variation of accretion flux and stellar parameters: non-LTE
calculations should be carried out, at least, for Fe, Mg, Ti and Na.  These
lines will be used in the future for Doppler mapping of CTTS.  For example
it has been shown in this work that Ca\,I at 4226.7{\AA} resonance line is
sensitive to spot's orientation relative an observer, while, at fixed
veiling, the line of He\,II at 4686{\AA} and the lines of Ca\,II are
sensitive to pre-shock gas density.

4) X-ray and UV spectra of the accretion shock should be calculated for the
parameters $\log N_0>13$ and $V_0>400$ km\,s$^{-1},$ i.e.  larger than used
by Lamzin (1998), because there are reasons to suppose that they may be
necessary to model spectr of some CTTS.

5) Lines of some important molecules, which have not been included yet in
our simulation of spectra, should be added, for instance, TiO, because the
molecular lines are typical for late type stars.  What is more in addition
to the hot spot, cold spots may also present on the surface of CTTS (Donati
et al., 2011).

\bigskip

 We wish to thank L.I. Mashonkina for useful discussions, for assistance in
testing the programs and for the atomic data for calcium; referees for
usefull notes and P.P.Petrov, who drew our attention to the problems
with interpretation of veiling of Ca\,I lines in spectra of CTTS. The 
paper based on observations made with ESO Telescopes at the La Silla Paranal
Observatory under programme ID 078.A-9059.  This research has made use of
the Keck Observatory Archive (KOA), which is operated by the W.M.  Keck
Observatory and the NASA Exoplanet Science Institute (NExScI), under
contract with the National Aeronautics and Space Administration.  Principal
investigators of the observations were P.\,Butler, G.\,Marcy, S.\,Vogt,
G.\,Herczeg, J.\,Johnson, A.\,Sargent, L.\,Hillenbrand.  The work was
supported by the Program for Support of Leading Scientific Schools
(NSh-5440.2012.2).


\section*{The Appendix. The model of the helium atom}

  The model of the helium atom, used in our calculations,  includes energy
levels of He\,I up to the principal quantum number $n=7$ with an orbital
splitting and components of the fine structure, energy levels of He\,II up
to $n=10$ without the orbital splitting and the state of He\,III.  Thus, 66
energy levels were included in our model, however at the calculation of the
level population, the levels of He\,I with $n=4-7$ and He\,II with $n=6-10$
were combined into the superlevels. A relatively small number of the levels
in the model is primarily caused by the lack of precise values of rate
coefficients for collisional excitation and ionization from upper levels. 
In other words, it is reasonable to assume that the increase in accuracy due
to a larger number of levels will be compensated by low quality of
the atomic data for these levels.

  Sublevels of He\,I and He\,II superlevels assumed to be populated
according to the Boltzmann distribution.  Due to this reason it was not
necessary to consider electron transitions within the superlevel. 
Transitions between the superlevel $S$ and other levels $n$ of the atom are
taken into account by effective transition rates $C,$ which were obtained by
averaging of transition rates $C_{Sp\rightarrow n}$ from a sublevel $Sp$ to
level $n$ over the Boltzmann distribution for the sublevels $Sp$. 
Transition rates from the levels $n$ to the sublevels $Sp$ of the superlevel
$S$ are calculated by a simple summation.  This approach is fully justified
at the high density in an accretion column: test calculation without the
combination of the levels leads to the same results and shows that the
upper levels are indeed populated according to the Boltzmann distribution.

  To get a sense of the influence of the number of levels, taken into
account in He\,I and He\,II, on the accuracy of the results, we used the
Cloudy08 code (Ferland et al., 1998).  As was noted in Sect.1 the Cloudy08
code and our calculations give similar results for a model of a homogeneous
slab with parameters similar to the parameters of the gas in the formation
region of He\,I and He\,II lines: $T=30\,000$~K, $N=10^{14.5}$ cm$^{-3},$
slab thickness $H=10^3$ cm, blackbody radiation field with $T=30\,000$~K and
a dilution factor $10^{-3}.$

  At first we have calculated, using the model of the slab, the optical
depth of a few lines of He\,I and He\,II in the case of the atomic model,
which included 7 levels for He\,I and 10 for He\,II.  Then we have repeated
the calculation, but taking 17 levels for He\,I and 20 for He\,II.  It
appeared that relative difference in optical depths of various lines of
He\,II with low level principal quantum number $n=1-4$ was about 0.5\%.  In
the case of He\,I the difference for $n=1-2$ is less than 1\%, and for $n=3$
it is about a few percents.  Analogous tests for the models with $N=10^{13}$
and $N=10^{15}$ cm$^{-3}$ have led to similar results and we concluded that
the number of the levels in our atomic model was sufficient to solve the
considered problem.


\end{document}